\documentclass{aa}  
\usepackage{graphicx}

\usepackage[varg]{txfonts}
\usepackage{tablefootnote}
\usepackage{subfigure}
\usepackage[symbol]{footmisc}

\usepackage{xcolor}

\usepackage{natbib}
\bibpunct{(}{)}{;}{a}{}{,} 

\usepackage{graphicx}
\usepackage[utf8]{inputenc}
\usepackage[english]{babel}
\usepackage{amsmath}
\usepackage{wasysym}
\usepackage{tabularx}
\usepackage{multirow}
\usepackage{booktabs}
\usepackage{amssymb}
\usepackage{rotating}
\usepackage{hyperref}
\makeatletter
\let\linenumbers\relax

\makeatother

\begin{document}

\title{CLASH-VLT velocity anisotropy profiles in a stack of massive galaxy clusters}
\author{Enrico Maraboli\inst{1}
  \and Andrea Biviano\inst{2,3}
  \and Claudio Grillo\inst{1,4}
     \and Amata Mercurio\inst{5,6}
     \and Lorenzo Pizzuti\inst{7}
     \and Piero Rosati\inst{8,9}
     \and Maurizio D'Addona\inst{5,6}
     }


\institute{Dipartimento di Fisica, Università degli Studi di Milano, Via Celoria 16, I-20133 Milano, Italy
  \and INAF $-$ Osservatorio Astronomico di Trieste, via G.B. Tiepolo 11, 34143 Trieste, Italy
  \and IFPU $-$ Institute for Fundamental Physics of the Universe, via Beirut 2, 34014 Trieste, Italy
  \and INAF $-$ IASF Milano, via A. Corti 12, I-20133 Milano, Italy
  \and INAF $-$ Osservatorio Astronomico di Capodimonte, Via Moiariello 16, 80131 Napoli, Italy
  \and Università di Salerno, Dipartimento di Fisica "E.R. Caianiello", Via Giovanni Paolo II 132, I-84084 Fisciano (SA), Italy
  \and Dipartimento di Fisica G. Occhialini, Università degli Studi di Milano Bicocca, Piazza della Scienza 3, I-20126 Milano, Italy
  \and Department of Physics and Earth Sciences, University of Ferrara, Via G. Saragat, 1, 44122 Ferrara, Italy
  \and INAF $-$ OAS, Osservatorio di Astrofisica e Scienza dello Spazio di Bologna, via Gobetti 93/3, I-40129 Bologna, Italy
}


\abstract{We measure the velocity anisotropy profile $\beta(r)$ of different galaxy cluster member populations by analysing the stacked projected phase space of nine massive ($M_\mathrm{200c}>7\times10^{14}$ M$_\odot$) galaxy clusters at intermediate redshifts ($0.18 < z < 0.45$). We select our sample of galaxy clusters by choosing the most round and virialised objects among the targets of the CLASH-VLT spectroscopic program, which offers a large spectral database. Complementary MUSE observations on most of these clusters allowed us to identify an unprecedented number of cluster members, strongly enhancing the precision of our measurement with respect to previous studies. Our sample of cluster members is divided in four classes: the first two are based on colour, red and blue galaxies, while the other two on stellar mass, high and low. To study the velocity anisotropy profile of each cluster member population, we employ two parallel techniques, namely the \texttt{MAMPOSSt} method (parametric in $\beta(r)$) and the inversion of the Jeans equation (non parametric in $\beta(r)$). The results from both techniques are found in agreement for any given cluster member population, and suggest that the orbital anisotropy in galaxy clusters grows from the centre (where $\beta\approx 0.2-0.4$) to the virial radius ($\beta\gtrsim 0.8$), and it is similar for the different cluster member populations. We also find an interesting dynamical feature emerging from the Jeans inversion results, that is a sudden drop in $\beta(r)$ at a distance of $\sim 250$ kpc from the cluster centre. We provide robust anisotropy estimates by exploring a highly significant number of model combinations: 72 with \texttt{MAMPOSSt} (varying the mass, surface number density, $\beta(r)$ model, and galaxy population) and 18 (varying total mass model and galaxy population) in the Jeans inversion.
Such an extensive investigation of the velocity anisotropy profile in galaxy clusters is a wide basis for future studies on cluster dynamical masses and cluster cosmology in the era of large spectroscopic surveys. }

\maketitle

\section{Introduction} \label{sec.intro}

Galaxy cluster dynamics is a powerful tool to investigate several fundamental properties of these structures, such as their evolution history and their total masses. Among the dynamical processes that shape cluster's assembly, the virialisation (or relaxation) is a key step which has long been investigated (e.g. \citealt{Chandrasekhar_1942_relaxation,LyndenBell_1967_violentrelax}). Relaxed clusters are indeed optimal candidates to undergo an astrophysical analysis, since they reached dynamical equilibrium and they are often approximately spherical. Under these hypotheses, many theoretical tools can be exploited to study these structures, and in this paper we make use of the Jeans equation (JE) of dynamical equilibrium (e.g. \citealt{GalacticDynamics}). A key quantity in this equation is the velocity anisotropy profile, $\beta(r)$, that is classically defined as

\begin{equation}
    \beta(r)= 1- \frac{\sigma_\vartheta^2(r) +\sigma_\varphi^2(r)}{2\sigma_r^2(r)}= 1- \frac{\sigma_\vartheta^2(r)}{\sigma_r^2(r)}, 
    \label{eq.beta}
\end{equation}
where $\sigma^2_r(r),\sigma_\vartheta^2(r)$, and $\sigma_\varphi^2(r)$ are the diagonal elements of the velocity dispersion tensor, and the second equality follows under the assumption of no streaming motions. The velocity anisotropy profile is tightly related to the shape of the orbits of cluster members: if $\beta=1$, orbits are purely radial; if $\beta\rightarrow -\infty$, orbits are circular; if $\beta=0$, orbits are isotropic. The Jeans equation establishes a direct connection between the total mass profile $M(r)$, the velocity dispersion profile along the radial direction, $\sigma_r(r)$, and the number density profile of the tracers of the gravitational potential, $\nu(r)$: 

\begin{equation} \label{eq.jeansmass}
    M(r)=-\frac{r\sigma^2_r(r)}{G}\left\{\frac{\mathrm{d}\ln \nu(r)}{\mathrm{d}\ln r} + \frac{\mathrm{d}\ln \sigma^2_r(r)}{\mathrm{d}\ln r }+2\beta(r)\right\}.
\end{equation}

It is hence clear that in order to make an accurate dynamical measurement of the total mass profile in galaxy clusters, as well as other dynamics-related quantities, a precise knowledge of the orbital velocity anisotropy is a matter of the highest priority. This will become more and more important with the huge amount of spectroscopic data that will come in the near future from all the current and next-generation facilities. 

In this work we will also adopt a different parametrisation of the velocity anisotropy, that is $\beta'(r)=\sigma_r/\sigma_\vartheta =1/\sqrt{1-\beta(r)}$.
This parametrisation offers better insights on the behaviour of the velocity anisotropy, specially when orbits are more radial, since, unlike $\beta(r)$, $\beta'(r)$ is not capped at 1. 

The Jeans equation has always been a key tool to study dynamically relaxed objects, and with the ever-growing availability of spectroscopic data (e.g. \citealt{Rosati_2014_CLASH,Treu_2015_GLASS,Bezanson_2024_UNCOVER}) it has been more and more exploited in galaxy cluster analyses to measure the velocity anisotropy profile. 

Another important constraining method for $\beta(r)$ is the analysis of the whole projected phase space (PPS, e.g. \citealt{Merritt_1987_ComaDM,Wojtak_2008_PPSdistribution,Wojtak_2009_PPSanisotropytest,Wojtak_2010_concmassrelation}), which in recent years has been many times performed with \texttt{MAMPOSSt} \citep{Mamon_2013_MAMPOSSt} and its extension to modified gravity theories \texttt{MG-MAMPOSSt} \citep{Pizzuti_2021_MGMAMPOSStphysics,Pizzuti_2023_MGMAMPOSSt}. \texttt{MAMPOSSt} performs a parametric fit of the total mass profile $M(r)$ and the velocity anisotropy profile $\beta(r)$ based on the surface density of the cluster member distribution in the PPS. Although a gaussian assumption on the 3D velocity distribution does not allow to recover the correct kurtosis at all projected radii (as already recognized in the original \texttt{MAMPOSSt} paper), \texttt{MAMPOSSt} were shown to produce unbiased and precise estimates of the velocity anisotropy profiles via several tests based on simulated halos (\citealt{Mamon_2013_MAMPOSSt,AguirreTagliaferro_2021_MAMPOSStvalidationNbodysim,Read_2021_MAMPOSStvalidationNbodysim}). Its vast applications in the literature \citep{Biviano_2013_M1206,Biviano_2016_highzanisotropyGCLASS,Biviano_2017_concentrationrelation,Biviano_2024_jellyfishorbits,Annunziatella_2016_A209speccat,Capasso_2019_SPTsurveyanisotropy,Mamon_2019_WINGSanisotropy,Sartoris_2020_AS1063,Pizzuti_2025_CHEXMATEanisotropy,Valk_2025_anisotropySDSSlowz} proved this method among the most tested and reliable ones to study velocity anisotropy profiles in galaxy clusters.

In this paper we focus our efforts on characterising $\beta(r)$ with an unprecedented precision through the resolution of the JE, exploiting the constraining power of the high number of cluster members offered by stacking multiple galaxy clusters. This method has been employed several times in the literature, because it allows to overcome the problem of small number statistics for individual clusters \citep{Biviano_2004_ENACSanisotropy,Biviano_2009_anisotropiesnearfar,Biviano_2016_highzanisotropyGCLASS} and therefore to enhance the statistical significance of the results \citep{Biviano_2017_concentrationrelation,Mamon_2019_WINGSanisotropy,Capasso_2019_SPTsurveyanisotropy,Pizzuti_2025_CHEXMATEanisotropy,Valk_2025_anisotropySDSSlowz}. The resulting stacked cluster exhibits the mean properties of the cluster sample.
\cite{Biviano_2025_CLASHanisotropisingle} has recently studied the $\beta(r)$ of the individual clusters we use in the stacking, but due to insufficient number statistics, they did not separate different cluster galaxy populations. Spotting peculiar features of the various galactic populations is truly crucial to understand the evolutionary path followed by the whole galaxy cluster throughout its life \citep{Butcher_1978_clustergalacticcontent,Dressler_1997_morphdensevolution}. The orbits of cluster members trace the dynamical evolution of their corresponding populations, and it is worth the effort to study each of them.

The stacking technique not only provides a statistics much better than available for individual clusters, but, in addition, it also allows to reduce the impact of systematics. In fact, the stacked cluster is by construction "more spherical" and with less evident internal substructures than its parent clusters, and then it is a better candidate to meet the assumptions of the spherical JE. 

The paper is organized as follows. In Section 2, we present in detail the adopted cluster sample and the photometric and spectroscopic data employed herein. In Section 3 we illustrate the whole cluster member selection process, alongside with the evaluation of the spectroscopic completeness of our sample, that allows us to build the projected number density profiles. The subsequent categorisation of the galaxies in our sample is then explained in Section 4. Then, the two methods of our analysis, namely the \texttt{MAMPOSSt} method and the Jeans inversion, are presented in Section 5. In Section 6, we show the outcomes of our analysis in all the tested configurations among the two methods. Finally, in Section 7, we thoroughly review all the relevant features of both our results and methodologies. 

Throughout the paper, we assume a flat $\Lambda$CDM cosmological model, in which the Hubble constant value is $H_0 = 70 \ \mathrm{km \, s} ^{-1} \ \mathrm{Mpc}^{-1}$ and the total matter density value is $\Omega_\mathrm{m}$ = 0.3. 

\section{Data sets} \label{sec.data}

We choose a sample of nine, massive ($M_{200\mathrm{c}}>0.75 \times 10^{15} \, \mathrm{M}_\odot$)\footnote{$M_{200\mathrm{c}}$ corresponds to the total mass within a sphere inside which the mean mass density is 200 times the value of the critical mass density of the Universe at the redshift of the cluster.} galaxy clusters at intermediate redshifts ($0.18<z\leq0.45$) from the Cluster Lensing And Supernova survey with Hubble CLASH \citep{Postman_2012_CLASH}, particularly among those belonging to the CLASH-VLT \citep{Rosati_2014_CLASH} spectroscopic follow-up programme at ESO VLT: Abell 383, Abell 209, RX J2129.7$+$0005, MS2137$-$2353, RXC J2248.7$-$4431, MACS J1115.9$+$0129, MACS J1931.8$-$2635, MACS J1206.2$-$0847, and MACS J0329.7$-$0211. In Table \ref{tab.clusters} we report the main features of the clusters studied herein, such as their redshifts, $M_{200\mathrm{c}}$ values, $c_{200\mathrm{c}}$ values, and number of selected members. Hereafter we will refer to them with their shortened names: A383, A209, R2129, MS2137, AS1063 (from Abell S1063, the other name for RXC J2248.7$-$4431), M1115, M1931, M1206, and M0329. These clusters are round and virialised, optimal candidates for the requirements of the techniques that involve the Jeans equation. The sphericity of the clusters is justified by the similar $R_\mathrm{200c}$ values obtained in the lensing analyses of \cite{Umetsu_2018_CLASHmassWL} and \cite{Umetsu_2014_CLASHmasses}, the former adopting an elliptical NFW profile, and the latter a spherical one.
Likewise, the virialised status is indicated by the fact that similar mass estimates are obtained from gravitational lensing and from kinematics \citep{Biviano_2025_CLASHanisotropisingle}. The virialised status of the cluster is, indeed, a fundamental requirement of the JE, and for our stack is inherited from the original sample.
Moreover, the top quality data that are available for these galaxy clusters allow us to use a very wide sample of cluster members, lowering the statistical errors and allowing insightful characterisation of their orbits. In this paper, we exploit the photometric dataset provided mostly by the 8.3-m Subaru Telescope and a mixed spectroscopic dataset (mainly by VIMOS and MUSE), as we explain in the following.

\begin{table}
\centering
\caption{List of the galaxy clusters studied in this paper.}
\label{tab.clusters}
\begin{tabular}{ccccc} 
\toprule
Cluster & $z$ & $M_{200\mathrm{c}}$ ($10^{15} \, \mathrm{M}_\odot$) & $c_{200\mathrm{c}}$ & Members \\
\midrule
A383 & 0.187 & $1.02 \pm 0.43$ & $2.5 \pm 1.6$ & 523 \\
A209 & 0.209 & $1.93 \pm 0.36$ & $3.4 \pm 0.7$ & 1001 \\
R2129 & 0.234 & $0.78 \pm 0.24$ & $2.9 \pm 1.2$ & 333 \\
MS2137& 0.313 & $1.08 \pm 0.32$ & $2.4 \pm 1.0$ & 371 \\
AS1063 & 0.348 & $1.98 \pm 0.60$ & $1.6 \pm 0.7$ & 1142 \\
M1115 & 0.352 & $1.79 \pm 0.38$ & $2.5 \pm 0.7$ & 638 \\
M1931 & 0.352 & $1.16 \pm 0.28$ & $7.8 \pm 1.7$ & 359 \\
M1206 & 0.439 & $1.51 \pm 0.32$ & $5.8 \pm 1.7$ & 543\\
M0329 & 0.450 & $1.27 \pm 0.22$ & $5.4 \pm 1.3$ & 438 \\
\bottomrule 
\end{tabular}
\tablefoot{For each cluster, the columns show from the first one on the left: 1) shortened name; 2) (mean) redshift; 3) $M_{200\mathrm{c}}$ values; 4) $c_{200\mathrm{c}}$ values; 5) Number of selected members. All $M_{200\mathrm{c}}$ and $c_{200\mathrm{c}}$ values come from \cite{Umetsu_2018_CLASHmassWL}.}
\end{table}

\subsection{Photometric data} \label{subsec.photo}
The photometric dataset that we employ in our work is based on the wide-field observations of the Suprime-Cam imager \citep{Miyazaki_2002_SuprimeCam} at the Subaru telescope, which covers a field of view of $34'\times 27'$. All the clusters in our sample were observed with this instrument, except for AS1063, being too southern for Subaru. Insted, we employ the AS1063 observations by \cite{Gruen_2013_AS1063wl} with the Wide-Field Imager at the ESO 2.2-m MPG/ESO telescope at La
Silla. In the work by \cite{Umetsu_2014_CLASHmasses}, in Section 4.2 (as well as in Tables 1 and 2), there is a detailed description of the available multiband images for each galaxy cluster considered by us. To summarise, in these observations every cluster was observed at least in four optical passbands (B, V, R, I), which exposures range from 1000 to 10000 s per passband. 

The processing of these ground-based photometric observations follows basically the procedure described in Appendix A1 of \cite{Mercurio_2021_AS1063spectrcatalog}, that we summarise hereafter. First, the SExtractor software \citep{Bertin_1996_SExtractor} for the identification of luminous sources is employed, together with the PSF deconvolution operated by the PSFEx software \citep{Bertin_2011_PSFEx}, to create the photometric catalogues of each considered galaxy cluster. In this paper we analyse B, V, R, I band images, and we make an independent catalogue for each of these wavebands, that is then matched with the other three through the STILTS library \citep{Taylor_2006_STILTS}. To build each catalogue, we adopt a two-phase method (employed also in \citealt{Mercurio_2015_photocatSSS}) to better identify all the sources in every field of view. This method consists in running SExtractor two times: the first (called \textit{cold mode}) focuses on properly deblending the brightest and extended sources; the second (called \textit{hot mode}), instead, focuses on finding faint objects and splitting close sources. At the end of the procedure, we merge the two catalogues produced by the cold and the hot mode, respectively, deleting multiple detections of extended sources and replacing these objects in the catalogue of sources detected in hot mode.

To distinguish galaxies from point-like sources, for each galaxy cluster we plot, exactly like in the upper-right corner of Figure A1 in \cite{Mercurio_2021_AS1063spectrcatalog}, the R-band Kron magnitude and the half-flux radius measured by SExtractor. This plot allows us to separate stars from the rest of the sample, since their half-flux radius is almost constant for all R-band magnitudes. At high R magnitudes, the locus occupied by the stars is no longer distinguishable from that occupied by the galaxies, but the magnitude cuts that we will present in the next section prevent any star contamination. Finally, the catalogues in all bands have been visually inspected on the images to check the residual presence of spurious or misclassified objects.

\subsection{Spectroscopic data} \label{subsec.spectro}

The main source of spectroscopic data for the galaxy clusters in this paper is the CLASH-VLT programme \citep{Rosati_2014_CLASH}, based on the VIsual Multi-Object Spectrograph (VIMOS; \citealt{LeFevre_2003_VIMOS}), which had a field of view of about 10 Mpc at the median redshift ($z\approx$0.4) of the cluster sample, and each galaxy cluster in our sample was covered with eight to twelve pointings, with a total area of $\approx 15 \times 20$ arcmin$^2$. Depending on the specific cluster, spectroscopic observations were made in MOS LR blue configuration, that has a spectral resolution of $\sim120$ over the spectral range 3700 - 6700 $\mathrm{\r{A}}$, and a spectral sampling of 5.3 $\mathrm{\r{A}}$/pix, or in MOS MR configuration, that has a spectral resolution of $\sim580$ over the spectral range 5000 - 10000 $\mathrm{\r{A}}$, and a spectral sampling of 2.5 $\mathrm{\r{A}}$/pix. The other main spectroscopic data source for the galaxy clusters in this paper is the Multi Unit Spectroscopic Explorer (MUSE; \citealt{Bacon_2010_MUSEinstrument}), that has a 1 arcmin$^2$ field of view, a spatial sampling of 0.2 arcsec, a spectral resolution of $\sim2.4$ $\mathrm{\r{A}}$ over the spectral range 4750 - 9350 $\mathrm{\r{A}}$, and a spectral sampling of 1.25 $\mathrm{\r{A}}$/pix. MUSE observations are available for every galaxy cluster studied in this paper, except for MS2137.

In total, our spectroscopic collection from the nine galaxy clusters contains 24100 high quality redshifts, with an average of more than 2650 redshifts per pointed cluster.

\section{Membership and completeness} \label{sec.speccomp}

Before our membership and completeness analyses, we introduce a R-band magnitude cut for each cluster photometric and spectroscopic catalog at R$=23$, except for A383 (R$=23.5$), AS1063 (R$=23.5$), and M1931 (R$=22.5$), to reduce the contamination of the cluster samples from background galaxies.

\subsection{Members selection} \label{subsec.membsel}

\begin{figure}
\vspace{-0.45cm}
\centering
\includegraphics[scale=0.39]{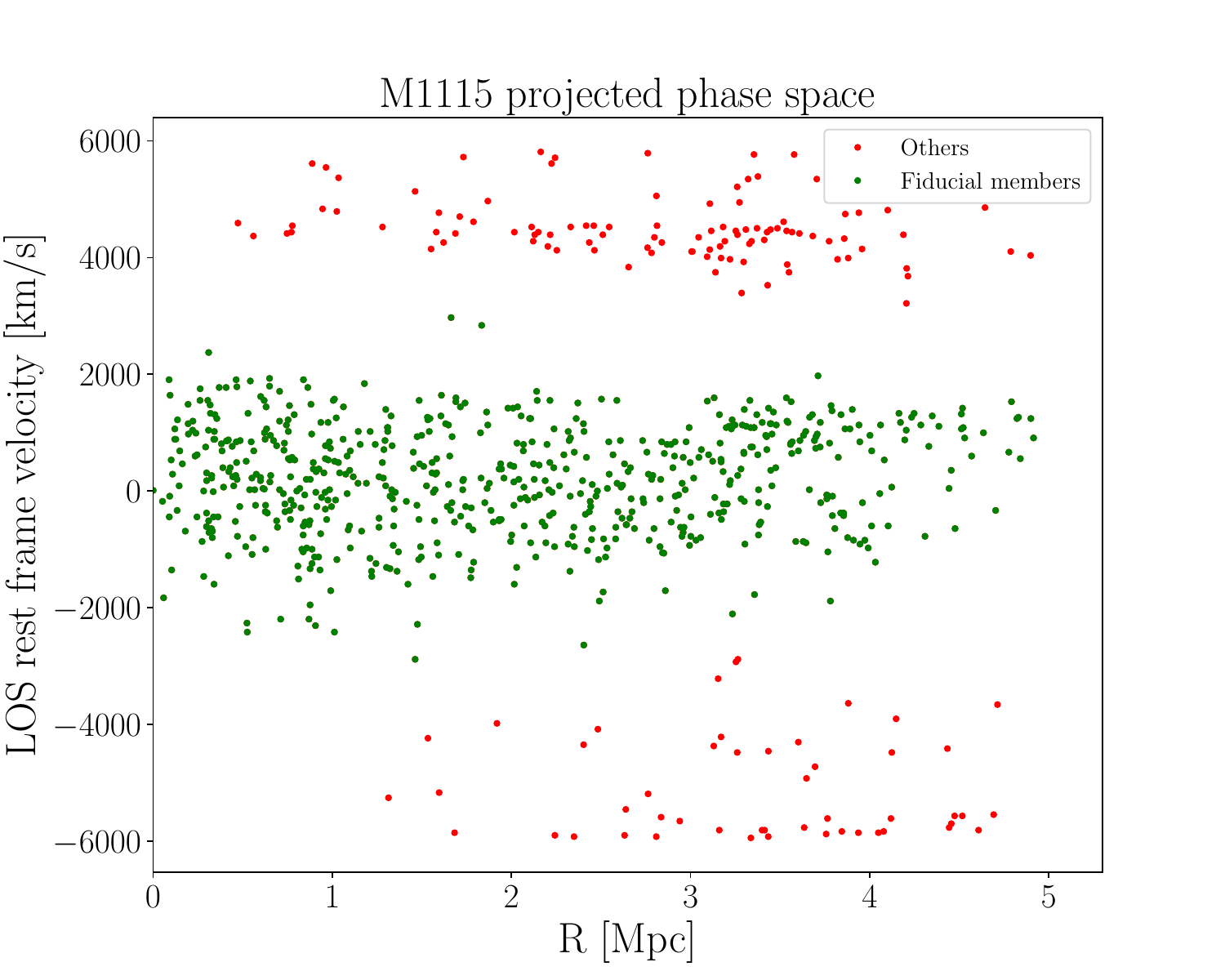}
\caption{Membership selection with CLUMPS on the PPS for the cluster M1115. The green dots represent the galaxy cluster members, while the red dots represent background and foreground galaxies.}
\label{fig.membership}
\vspace{-0.4cm}
\end{figure}

We base our choice of the cluster members on the \texttt{CLUMPS} algorithm (see Appendix A in \citealt{Biviano_2021_GOGREEN&CLUMPS} and also \citealt{Biviano_2025_CLASHanisotropisingle}). In summary, \texttt{CLUMPS} looks for local minima of the surface number density of spectroscopic objects in the PPS, with no assumption about the internal dynamics of the cluster. In this paper, we employ a modified version of \texttt{CLUMPS}, that returns a membership probability $P_\mathrm{mem}$ for each spectroscopic object, with recalibrated parameters $\mathrm{d}R$ and $\mathrm{d}V$. We calibrate the $\mathrm{d}R$ and $\mathrm{d}V$ parameters of CLUMPS on a sample of halos of similar mass and richness as our clusters from the GAlaxy Evolution and Assembly semi-analytical model (\texttt{GAEA}; \citealt{DeLucia_2007_GAEA, DeLucia_2024_GAEAclusters}), which is based on the Millennium simulation \citep{Springel_2005_Millenniumsimulation}. We adopt $\mathrm{d}R=450$ kpc and $\mathrm{d}V=200$ km/s that allows us to reach a purity $\geq 85$\% and a completeness $\geq 99$\% for galaxies with a CLUMPS membership probability $\geq 0.5$.  
In Figure \ref{fig.membership} we report the results of the membership selection for M1115 as an example, and we perform this operation separately for all the clusters. 

\subsection{Spectroscopic completeness and number density profile} \label{subsec.speccomp}

To compute the projected number density of the stacked cluster members $N(R)$, we have to rescale the projected distances of the cluster members from their corresponding cluster centre. We rescale each distance in units of the corresponding $R_\mathrm{200c}$ of the cluster, so that we can compare member distances of different clusters and merge them together to compute the $N(R)$ profile.
The evaluation of $N(R)$ can be affected by completeness issues of the spectroscopic data set of each individual cluster. Unaccounting for spectroscopic incompleteness could lead to an erroneous estimate of the intrinsic $\nu(r)$, if the completeness function is radial-dependent. On the other hand, the velocity distribution is not affected by completeness issues, since the observational selection does not operate in redshift space within the narrow redshift range spanned by each cluster. 

To determine the radial completeness of the cluster members only, we apply two colour cuts (described in detail in Appendix \ref{app.colorcuts}) to both our full photometric and spectroscopic samples in order to limit interloper contributions. The cutting process, as well as the whole spectroscopic completeness determination routine, is applied separately to every galaxy cluster in the sample. We assume as the centre of each galaxy cluster the position of the corresponding Brightest Cluster Galaxy (BCG hereafter), since our sample contains only round and virialised galaxy clusters with a single BCG. We divide in 10 bins, linearly spaced, the region within 0 and 600 kpc, while we divide in 5 bins, logarithmically spaced, the outer region from 600 to 6000 kpc. The aim of this spacing is to better sample  the inner, more populated, regions of the considered galaxy clusters. For each radial bin, we compute the ratio between the number of spectroscopic objects, members and non members, and the number of photometric objects that are within the considered bin after the two colour cuts. This ratio is the spectroscopic completeness of our sample. Hence, we assign to every cluster member the corresponding completeness coefficient, namely $C_i$, according to their radial distance from the BCG.

After this, we define another set of 100 bins spanning linearly from 0 to 6000 kpc, and for the $i$-th bin we compute the projected number density of members through the weighted sum of cluster members:

\begin{equation}
    N (\Bar{R_i})= \frac{1}{\pi(R_{i+1}^2-R_i^2)}\sum_\mathrm{bin}\frac{1}{C_i},
    \label{eq.surfdens}
\end{equation}
where $\Bar{R_i}=(R_{i+1}+R_i)/2$, and $R_i$ is the rescaled radial boundary of the considered bin. The uncertainty $\varepsilon(R)$ that we assign to the $N(R)$ profile is that associated to poissonian errors over counts, that is 

\begin{equation}
    \varepsilon (\Bar{R_i})= \frac{1}{\pi(R_{i+1}^2-R_i^2)}\sqrt{\sum_\mathrm{bin}\frac{1}{C_i}}.
    \label{eq.errsurfdens}
\end{equation}
To minimise its effect on the cluster dynamical state, we restrict our dynamical analysis to the virial region, that we define as the inner $1.2 R_\mathrm{200c}$ region, corresponding to radii $\leq R_\mathrm{100c}$. 

\section{Colour and mass subsampling}\label{sec.colormass}

\begin{table}
\caption{Number of spectroscopically confirmed cluster members for each galaxy cluster.  }
\label{tab.membersep}
\centering
\begin{tabular}{cccccc}
\toprule
Cluster & Red, HM & Blue, HM & Red, LM & Blue, LM \\
\midrule
A383 & 135 & 104 & 151 & 174 \\
A209 & 320 & 155 & 320 & 206 \\
R2129 & 157 & 36 & 94 & 34 \\
MS2137 & 128 & 70 & 58 & 115 \\
AS1063 & 425 & 159 & 322 & 236 \\
M1115 & 254 & 96 & 164 & 124 \\
M1931 & 218 & 66 & 42 & 33 \\
M1206 & 237 & 70 & 141 & 95 \\
M0329 & 183 & 168 & 21 & 66 \\
\midrule
Total & 2057 & 924 & 1313 & 1083 \\

\bottomrule
\end{tabular}
\tablefoot{Number of selected cluster members in the different sub-classes, where HM and LM stand for High (stellar) Mass and Low (stellar) Mass, respectively. The total number of cluster members is 5377, of which: 3370 red, 2007 blue, 2981 HM, and 2396 LM galaxies.}
\vspace{-0.4cm}
\end{table}

The core of this work consists in studying the velocity anisotropy profile of different populations of cluster members. In this paper we focus on four classes of galaxy cluster members: the first two are distinguished by colour, red and blue galaxies, while the other two by mass, high (HM) and low (LM). In Table \ref{tab.membersep} we report for each considered galaxy cluster the corresponding population subdivision, while in the next subsections we explain in detail how we assigned every cluster member in each category.

\subsection{Colour subdivision}\label{subsec.colorsub}

\begin{figure*}
\centering
\includegraphics[width=\textwidth]{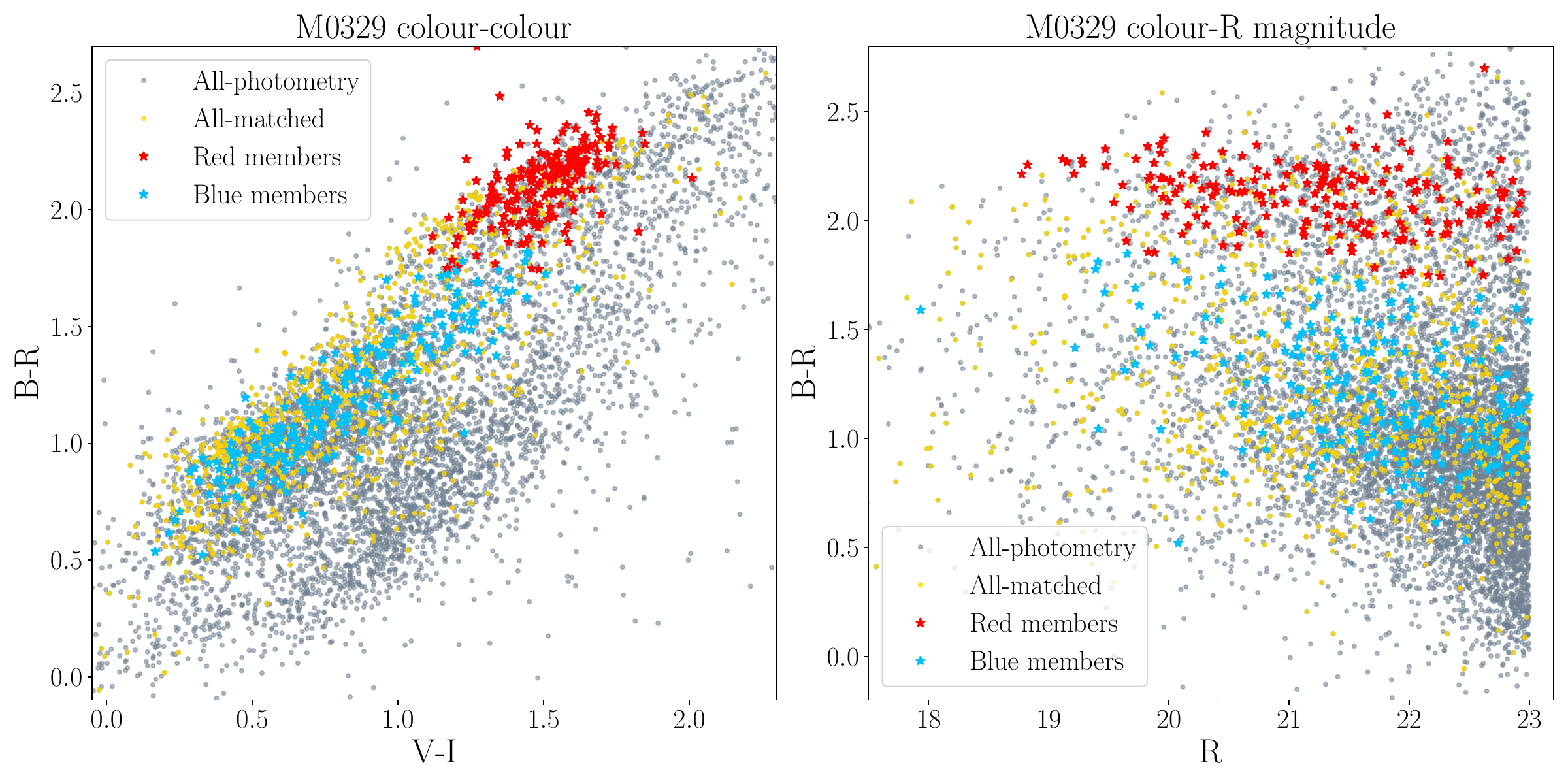}
\caption{The colour subdivision for the members of M0329, both on the colour-colour plane (V$-$I vs B$-$R, left panel) and on the R-band magnitude vs B$-$R colour plane (right panel). The latter is the plot used to determine which galaxies are red and which are not, according to the recursive red sequence linear fit. The "All-matched" entry in the legend, corresponding to the yellow points, represents the objects in the photometric catalogue that have a match in the spectroscopic catalogue.}
\label{fig.colorsep}
\end{figure*}

We select the red and blue members through a recursive linear fit of each cluster's red sequence in the R band magnitude versus B$-$R colour plot. In the $i+1$-th iteration of the linear fit, we exclude those cluster members which fall outside the 2 or 1 $\sigma$ region (depending on the cluster case) around the best fit of the $i$-th iteration. The algorithm stops when the $m_{i+1}$ angular coefficient of the linear best fit satisfies $(m_{i+1}-m_i)/m_i<0.03$. In Figure \ref{fig.colorsep} we show an example of colour subdivision for M0329, in which the red dots mark the cluster members belonging to the red sequence. Member galaxies in the so-found red sequence are classified as \textit{red}, while the others are labelled as \textit{blue}.

\subsection{Mass subdivision} \label{subsec.masssub}

The second subdivision we choose to do is related to the stellar mass $M_*$ of the selected cluster members. We divide our sample in two classes of galaxies: high and low mass (HM and LM hereafter) members. For the purposes of this simple subdivision, we adopt as a stellar mass proxy the I-band magnitude of cluster members (see e.g. \citealt{Pozzetti_2007_Ibandmassproxy,Mercurio_2021_AS1063spectrcatalog}) as it allows an easy identification of the two categories.
We separate the two components by identifying a dip in the I-band magnitude distribution of each cluster, at magnitude $I_\mathrm{dip}$, as it represents the crossing point of two different galactic mass functions \citep{Annunziatella_2016_A209speccat,Mercurio_2021_AS1063spectrcatalog}. 
Hence, we define as HM cluster members those with I$<I_\mathrm{dip}$ and, viceversa, as LM cluster members those with I$>I_\mathrm{dip}$. 

\section{Methods of the dynamical analysis} \label{sec.methods}

\begin{figure*}
\centering 
\includegraphics[width=1.05\textwidth]{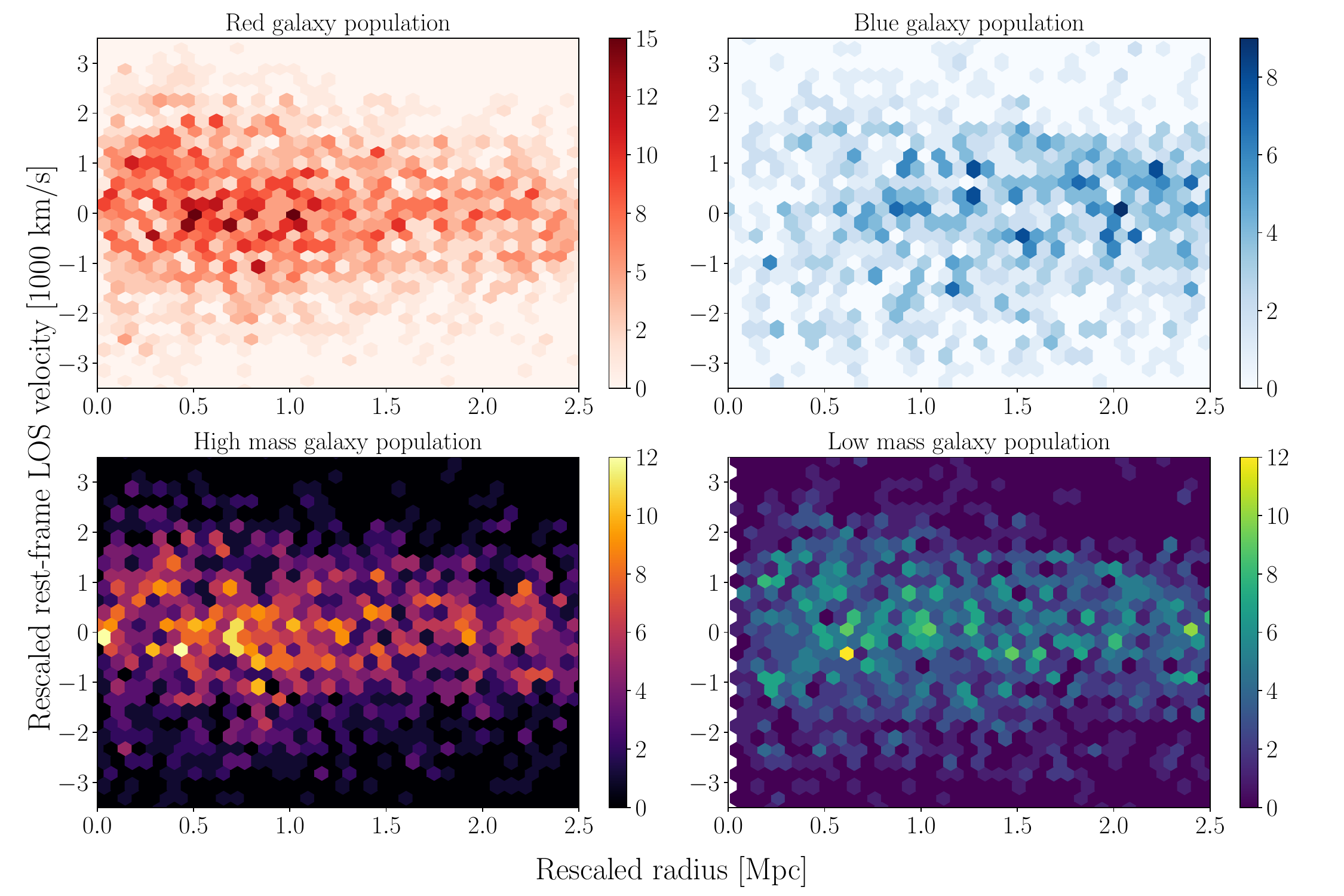}
\caption{Projected-phase space of the major cluster member populations in this paper, rescaled by the mean value of $R_\mathrm{200c}$ and $v_\mathrm{200}$.. The scale at the right of every plot represents the number of objects in each bin. We note that the blue galaxies, as expected, are more concentrated towards the outer regions of our ensemble cluster, while the red ones, viceversa, are mostly found in the inner regions. The high mass population instead, despite being radially distributed almost as the low mass population, is slightly less spread in velocity than the latter.}
\label{fig.galpopdistr}
\end{figure*}

According to a common practice in the existing literature, we stack the cluster members of our sample in a normalized PPS, under the hypothesis of homology between galaxy clusters \citep{Navarro_1996_concmassrelation,NFW_1997_DMprofile, Navarro_2004_homology}. We normalize the clustercentric distance of each member galaxy with respect to the original cluster $R_\mathrm{200c}$ value, and their rest-frame line of sight velocity with respect to the quantity

\begin{equation}
    v_{200}=\sqrt{\frac{GM_\mathrm{200c}}{R_\mathrm{200c}}}.
\end{equation}
We then rescale the normalized clustercentric radii and rest-frame line-of-sight (LOS, hereafter) velocities by the mean value of $R_\mathrm{200c}$ and $v_\mathrm{200}$, respectively, to emulate a real cluster. The newly formed stacked cluster is composed by 4176 cluster members, which belong to different populations: in Figure \ref{fig.galpopdistr} we present how different galaxy populations are distributed within the rescaled PPS.

In this work, we implement two different strategies to measure the velocity anisotropy profile $\beta(r)$ in the stacked galaxy cluster: the first approach to the dynamical problem is through a parametric description of the $\beta(r)$ profile, that is realized through the \texttt{MAMPOSSt} \citep{Mamon_2013_MAMPOSSt} method and the \texttt{MG-MAMPOSSt} \citep{Pizzuti_2021_MGMAMPOSStphysics,Pizzuti_2023_MGMAMPOSSt} code; the second approach is through the inversion of the Jeans equation (see, e.g., \citealt{BinneyMamon_1982_Jeansinv, Solanes_1990_S3Jeansinv, Dejonghe_1992_DMjeansinv} or Appendices A and B in \citealt{Mamon_2019_WINGSanisotropy} for equivalences between different inversion methods), that aims to compute $\beta(r)$ in a non parametric way. Employing two different methods also allows us to cross-check the results that we show in Section \ref{sec.results}.

\subsection{Parametric analysis with MAMPOSSt} \label{subsec.mamposst}
As mentioned in the Introduction, \texttt{MAMPOSSt} adopts parametric profiles for $\nu(r)$, $M(r)$, and $\beta(r)$. Throughout this paper we explore different combinations of the following models. 

For the number density $\nu(r)$, we test two well-known models:
\begin{itemize}
    \item the Navarro-Frenk-White (NFW) profile \citep{NFW_1997_DMprofile}, given by
    \begin{equation}
        \nu_\mathrm{NFW}(r)=\frac{\nu_\mathrm{0, NFW}}{\frac{r}{r_\mathrm{NFW}}\left(1+\frac{r}{r_\mathrm{NFW}}\right)^2} \, ;
        \label{eq.NFW}
    \end{equation}
    \item the Hernquist profile \citep{Hernquist_1990_profile}, given by 
    \begin{equation}
        \nu_\mathrm{H}(r)=\frac{\nu_\mathrm{0, H}}{4\pi} \frac{1}{\frac{r}{r_\mathrm{H}}\left(1+\frac{r}{r_\mathrm{H}}\right)^3}.
        \label{eq.hernq}
    \end{equation}
\end{itemize}
We directly fit the projected number density profile $N(R)$ with both the NFW and the Hernquist profiles, through a maximum likelihood bayesian algorithm (see \citealt{Maraboli_2025_virialquantitiesfromSL}), and we give the resulting parameter values as an input to \texttt{MG-MAMPOSSt}. In Figure \ref{fig.Nprofiles} we show the projected number density profiles $N(R)$ for the different galaxy populations studied in this work, and the corresponding NFW/Hernquist profile fits.

For the total mass profile $M(r)$ we employ again the NFW model, that is
\begin{equation}
    M_\mathrm{NFW}(r) = 4\pi\rho_\mathrm{NFW}r_\mathrm{NFW}^3\left[\ln{\left(1+\frac{r}{r_\mathrm{NFW}}\right)}- \frac{r}{r+r_\mathrm{NFW}}\right],
    \label{eq.nfwmass}
\end{equation}
and the Hernquist model,
\begin{equation}
    M_\mathrm{H}(r)=\frac{\rho_\mathrm{H}r_\mathrm{H}^3}{2} \frac{r^2}{(r_\mathrm{H}+r)^2}.
    \label{eq.hernmass}
\end{equation}

Finally, for the velocity anisotropy profile $\beta(r)$ we test two models from a general class of models, written as
\begin{equation}
    \beta(r)=\beta_0 + (\beta_\infty-\beta_0)\frac{r^\delta}{r^\delta+r_\beta^\delta}.
    \label{eq.betageneral}
\end{equation}
Here we have four free parameters: $\beta_0=\beta(r=0)$, $\beta_\infty=\lim_{r\rightarrow + \infty}\beta(r)$, $r_\beta$ is a scale radius and $\delta$ is a fixed exponent. In this paper we test the specific cases in which $\delta=1$, that is the known generalised Tiret (gT) model \citep{Tiret_2007_betamodel}, and in which $\delta=2$, known as the generalised Osipkov-Merritt (gOM) model \citep{Osipkov_1979_OMbetamodel,Merritt_1985_OMbetamodel}. For these models, \texttt{MAMPOSSt} assumes that $r_\beta=r_{-2}=r_\mathrm{NFW}=r_\mathrm{H}/2$, where $r_{-2}$ is the radius at which $\mathrm{d}\ln\rho/\mathrm{d}\ln r =-2$ and $\rho(r)$ is the total mass density profile.

We remark that in our analysis we take into consideration all the possible combinations of the $\nu(r)$, $M(r)$, and $\beta(r)$ profiles.

\subsection{Numeric Jeans inversion} \label{subsec.jeansinv}
The second method that we employ is the so-called Jeans inversion (JI), which is 
an almost fully non-parametric solution of the Jeans equation that allows to extract the $\beta(r)$ profile (\citealt{BinneyMamon_1982_Jeansinv,Solanes_1990_S3Jeansinv,Dejonghe_1992_DMjeansinv}). This technique, indeed, does not need to fit neither the stacked cluster's $N(R)$ nor its $\sigma_\mathrm{los}(R)$ profiles, which we just smooth with the LOWESS method (see, e.g., \citealt{Gebhardt_1994_GCcoressmoothingref}), and it only needs the total mass profile as an input. 

Our code for the Jeans inversion solves the Jeans equation in the formulation of both \cite{Solanes_1990_S3Jeansinv} and \cite{Dejonghe_1992_DMjeansinv}, which splits the problem in a set of simpler equations.  Since these equations contain integrals up to infinity, we have to extrapolate $N(R)$ and $\sigma_\mathrm{los}(R)$ to a sufficiently large radius that can mimic infinity. We choose to set this "infinity radius" $R_\mathrm{inf}$ at 30 Mpc from the stacked cluster centre, and in Section \ref{sec.discussion} we discuss thoroughly how the choice of this and other arbitrary parameters affects the results of the Jeans inversion. 

\subsubsection{Extrapolation of $N(R)$ and $\sigma_\mathrm{los}(R)$} \label{subsec.extrapolation}

The extrapolation process follows that described in \cite{Biviano_2013_M1206}, in which the $N(R)$ and the $\sigma_\mathrm{los}(R)$ profiles are treated separately. Beyond the last observed radius $R_\mathrm{last}$, for $N(R)$ the extrapolation is done according to Eq. 10 in \cite{Biviano_2013_M1206}, that, for completeness, we report here:
\begin{equation}
\label{eq.Nextrap}
    N(R>R_\mathrm{last})=\eta\frac{(R_\mathrm{inf}-R)^\xi}{R^\zeta},
\end{equation}
where
\begin{equation}
\label{eq.zetaextrap}
    \zeta=\left. \frac{\mathrm{d} \log N}{\mathrm{d} \log R}\right\vert_{R_\mathrm{last}}- \xi \frac{R_\mathrm{last}}{R_\mathrm{inf}-R_\mathrm{last}},
\end{equation}
and
\begin{equation}
\label{eq.etaextrap}
    \eta=N(R_\mathrm{last})\frac{R_\mathrm{last}^\zeta}{(R_\mathrm{inf}-R_\mathrm{last})^\xi}.
\end{equation}
We note that the only free parameter in this formulation is $\xi$. For our analysis, we set $\xi=0.5$ based on our experience, and we will discuss in Section \ref{sec.discussion} also the impact of this (and of $R_\mathrm{inf}$) choice on the results. The extrapolation method of $\sigma_\mathrm{los}(R)$, instead, assumes that $\sigma_\mathrm{los}(R_\mathrm{inf})=f\sigma_\mathrm{max}$, where $\sigma_\mathrm{max}$ is the $\sigma_\mathrm{los}(R)$ peak value and $f$ a fixed constant, and it is a log-linear interpolation:
\begin{equation}
\label{eq.sigmaextrap}
    \sigma_\mathrm{los}(R>R_\mathrm{last})=(f\sigma_\mathrm{max}-\sigma_\mathrm{last})\frac{\log R -\log R_\mathrm{last}}{\log R_\mathrm{inf} -\log R_\mathrm{last}}+\sigma_\mathrm{last},
\end{equation}
where $\sigma_\mathrm{last}=\sigma_\mathrm{los}(R_\mathrm{last})$. We set the value of $f$ to 0.2, since with this value we observe that $\sigma_\mathrm{los}(R_\mathrm{inf})$ assumes values comparable to the velocity dispersion of field galaxies. Once $N(R)$ and $\sigma_\mathrm{los}$ have been extrapolated, they can be employed in the JI equations of \cite{Solanes_1990_S3Jeansinv} and \cite{Dejonghe_1992_DMjeansinv}.

\subsubsection{Ensemble mass profile} \label{subsec.masscraft}

The last input required to obtain the $\beta(r)$ anisotropy profile from the Jeans inversion equations is the total mass profile of the stacked cluster. Following a common method in the literature \citep{Biviano_2004_ENACSanisotropy,Katgert_2004_MLratioENACS,Biviano_2009_anisotropiesnearfar,Mamon_2019_WINGSanisotropy}, we build a mass profile that mimics the average properties of the real galaxy clusters in our sample. We base this process on three fundamental considerations:
\begin{itemize}
    \item [a)] the projected numerical density $N(R)$ of the stacked cluster is the mean of each cluster $N(R)$, naturally weighted by their respective number of cluster members;
    \item [b)] the average of more analytic mass profiles, such as NFW or Hernquist, is not an analytic mass profile, whereas an analytic mass profile with average parameter values is;
    \item [c)] the physical scales of a cluster, such as its $R_\mathrm{200c}$ or $M_\mathrm{200c}$ are the most important ones to reproduce through the averaging process.
     
\end{itemize}
The first key point leads us to perform every average using the number of cluster members as weights. Hence, within this section the mentioned averages are always weighted on the number of members of the respective galaxy cluster, and written as $\langle \cdot \rangle$. The second key point makes us choose to compute the total mass profile via averaged parameters instead of averaging different total mass profiles. Doing so we can link the behaviour of the velocity anisotropy profiles that we show in Section \ref{sec.results} to a well-identified total mass profile. The last key point is directly linked to the computation of the total mass profile, where we refer to Eq. 11 in \cite{Mamon_2019_WINGSanisotropy} and its reformulation in terms of virial quantities:
\begin{equation}
    M(r)=M_\mathrm{200c}\frac{\Tilde{M}(c_\mathrm{200c}r/R_\mathrm{200c})}{\Tilde{M}(c_\mathrm{200c})}, 
    \label{eq.massprofgen}
\end{equation}
where
\begin{equation}
    \Tilde{M}_\mathrm{NFW}(x)=\frac{\ln(x+1)-x/(x+1)}{\ln 2 -1/2} \,; \ \ \ \Tilde{M}_\mathrm{H}(x)=9\left(\frac{x}{x+2}\right)^2
    \label{eq.adim}
\end{equation}
are the expressions for the NFW and Henrquist model, respectively.
Hence the quantities needed to compute these profiles are two to be chosen between $R_\mathrm{200c}$, $c_\mathrm{200c}$, and $r_{-2}=R_\mathrm{200c}/c_\mathrm{200c}$. From \cite{Umetsu_2018_CLASHmassWL} we obtain for every galaxy cluster in our sample the corresponding $R_\mathrm{200c}$, $c_\mathrm{200c}$, and $r_{-2}\equiv r_\mathrm{NFW}$. We set as $R_\mathrm{200c}$ of the ensemble cluster the value $\langle R_\mathrm{200c} \rangle =2.13 \pm 0.20$ Mpc, from which follows the corresponding $M_\mathrm{200c}$ value of $(1.51 \pm 0.41)\times 10^{15} \, \mathrm{M_\odot}$ (assuming $\langle z \rangle=0.315$ as the redshift of the stacked cluster). Since $\langle r_{-2} \rangle \neq \langle R_\mathrm{200c} \rangle/\langle c_\mathrm{200c} \rangle$ and so on with the other relations between $R_\mathrm{200c}$, $c_\mathrm{200c}$, and $r_{-2}$, following key point c) we define the average concentration of the "synthetic" mass profile as 
\begin{equation}
    \Bar{c}_\mathrm{200c}=\frac{\langle R_\mathrm{200c} \rangle}{\langle r_{-2} \rangle}.
    \label{eq.concentration}
\end{equation}
We have to remark that the data from \cite{Umetsu_2018_CLASHmassWL} about $M_\mathrm{200c}$ and $c_\mathrm{200c}$ are obtained through a NFW fit of the cosmic shear measurements, and it is worth to check their robustness when transposed to a Hernquist model. Naturally, in our sample of round and virialised galaxy clusters each of them should have a single value of $r_{-2}$ of their mass density distribution, that can be consequently identified as $r_{-2}\equiv r_\mathrm{NFW} \equiv r_\mathrm{H}/2$. However, the $r_{-2}$ predicted from the NFW fit can be shifted by some systematics in the analysis with respect to a hypothetical Hernquist fit of the cosmic shear measurements. Hence, we test the "single $r_{-2}$ hypothesis" by exploiting the results of \cite{Maraboli_2025_virialquantitiesfromSL}, in which the authors fit the total mass profile of each cluster in their sample with different mass models, including NFW and Hernquist. We find that for all the clusters in \cite{Maraboli_2025_virialquantitiesfromSL} the relative discrepancy between $r_{-2}^\mathrm{NFW}$ and $r_{-2}^\mathrm{H}$ is $10\% < (r_{-2}^\mathrm{NFW}-r_{-2}^\mathrm{H})/r_{-2}^\mathrm{NFW}=\delta < 15\%$. Then, from \cite{Umetsu_2018_CLASHmassWL} we can obtain directly $r_{-2}^\mathrm{NFW}$ for each galaxy cluster in our sample, compute $\Bar{c}_\mathrm{200c}^\mathrm{NFW}=\langle R_\mathrm{200c} \rangle/\langle r_{-2}^\mathrm{NFW} \rangle$ and repeat the process for $\Bar{c}_\mathrm{200c}^\mathrm{H}=\langle R_\mathrm{200c} \rangle/\langle r_{-2}^\mathrm{H} \rangle$, where the $r_{-2}^\mathrm{H}$ of each cluster is computed as $r_{-2}^\mathrm{H}=(1-\delta)r_{-2}^\mathrm{NFW}$. With this recipe for the total mass profiles, as we do in the \texttt{MAMPOSSt} analysis, also with the Jeans inversion we study two different cases of them: the NFW and the Hernquist profile. For these two models, we compute the following scale values according to our method: $r_\mathrm{NFW}=0.81 \pm 0.44$ Mpc, $\Bar{c}_\mathrm{200c}^\mathrm{NFW}=2.7 \pm 1.5$, $r_\mathrm{H}=1.43 \pm 0.8$ Mpc, $\Bar{c}_\mathrm{200c}^\mathrm{H}=3 \pm 1.7$.

\section{Results} \label{sec.results}

\begin{figure*}
    \centering
    \includegraphics[width=\textwidth]{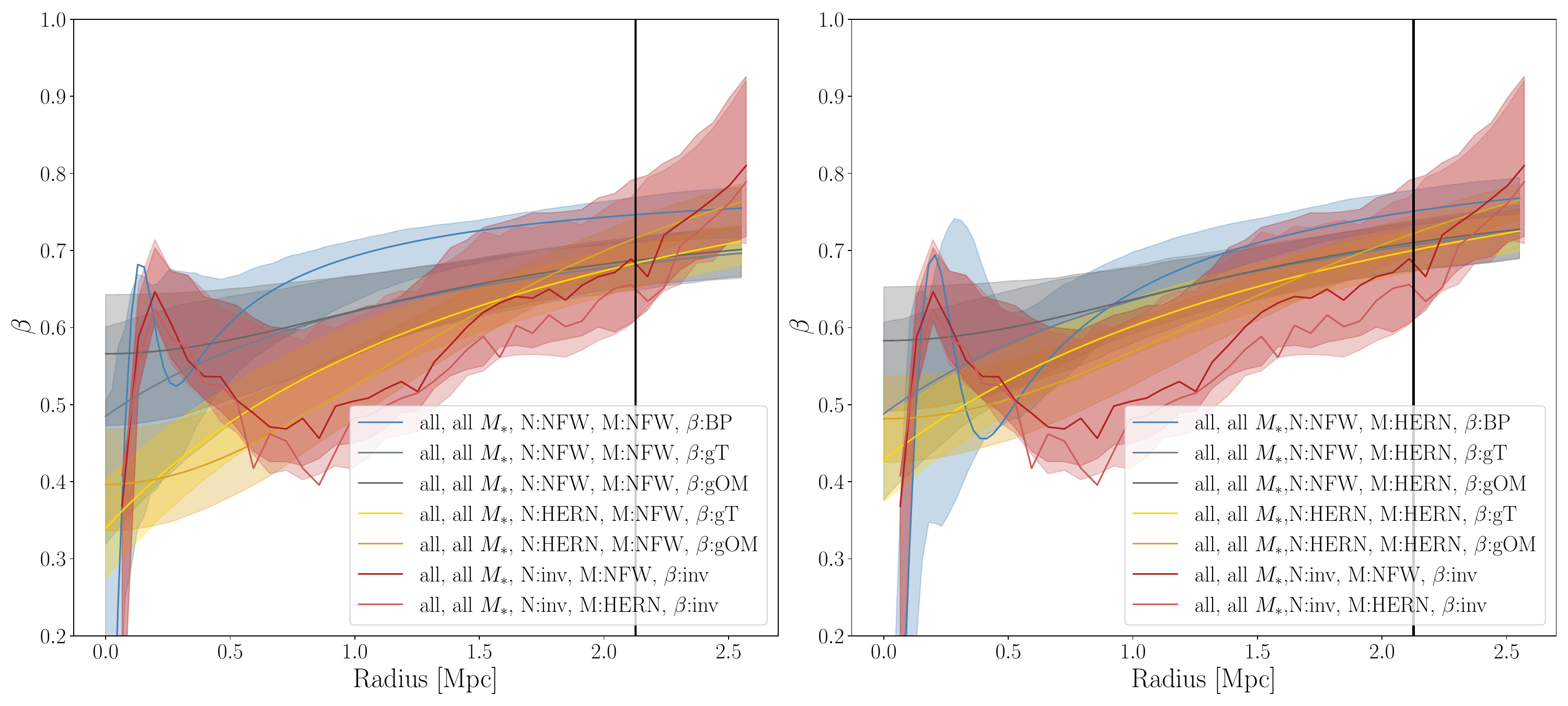}
    \caption{Comparison between the $\beta(r)$ profiles from the two methods of the whole cluster member population. In the left panel, we plot all the \texttt{MAMPOSSt} outcomes with a NFW total mass profile, as well as both the JI results (NFW and Hernquist) for reference. In the right panel, similarly, we plot all the \texttt{MAMPOSSt} outcomes with a Hernquist total mass profile. The vertical black line represents the $R_\mathrm{200c}$ of the stacked cluster.}
    \label{fig.comparemodel}
\end{figure*}

Throughout our analysis we test a great number of different combinations of models (for mass, projected number density and velocity anisotropy profiles) and galaxy population. In the whole \texttt{MAMPOSSt} analysis, indeed, we explore all the possible configurations between 2 total mass models (NFW and Hernquist), 2 $N(R)$ models (again, NFW and Hernquist), 2 $\beta(r)$ models (gT and gOM), 3 colour population choices (red/blue population and all together), and 3 mass population choices (high/low mass population and all together), for a total of 72 combinations. On the other hand, in the Jeans inversion analysis the inputs are just the 2 choices of the total mass profile, the 3 colour population choices, and the 3 mass population choices, for a total of 18 combinations.

The confidence intervals that we report for the \texttt{MAMPOSSt} analysis are defined as follows: we extract 500 samples of parameters from the Markov chains produced by \texttt{MG-MAMPOSSt}, which generate 500 new $\beta(r)$ profile; then, at any radius $r_0$, we take the 16th and the 84th percentile of the $\beta(r_0)$ value distribution, and we consider them as the boundaries of our confidence levels. The best-fit profile is the one generated by the median value of the fitted parameters. For the confidence intervals of the Jeans inversion, we follow a method similar to the previous one, that we explain in full details in Section \ref{subsec.jeanserrs}. The best JI profile is computed as described in Section \ref{subsec.jeansinv} with all the parameter values that we list there. 

In Figure \ref{fig.comparemodel}, we show both the \texttt{MAMPOSSt} and the JI results for the whole population of cluster members. We added to this comparison a new anisotropy profile, indicated as BP \citep{Pizzuti_2025_BPbetamodel}, discussed in detail in section \ref{subsec.comparison}.

\subsection{MAMPOSSt results} \label{subsec.MAMPOSStres}

The $\beta(r)$ and $\beta'(r)$ profiles obtained from the \texttt{MAMPOSSt} analysis, shown in Figure \ref{fig.MAMmajorgOM} and \ref{fig.MAMmajorgT}, outline two major features:
\begin{itemize}
    \item The choice of total mass and projected number density model can impact the resulting velocity anisotropy profiles;
    \item In the vast majority of model configurations, there are no significant changes in the $\beta(r)$ and $\beta'(r)$ shapes for the different cluster members.
\end{itemize}
We plot all the \texttt{MG-MAMPOSSt} configuration results in Appendix \ref{app.betaplots}. Moreover, we report in Tables \ref{tab.betamajor} and \ref{tab.betaminor} the median values of $\beta_0$ and $\beta_\infty$ for each configuration of mass, number density, anisotropy model and galaxy population.

It is easy to verify that the anisotropy profiles computed with a generalised Tiret model (Figure \ref{fig.MAMmajorgT}) are strongly consistent with the corresponding anisotropy profiles computed with a generalised Osipkov-Merrit model (Figure \ref{fig.MAMmajorgOM}, see also Figure \ref{fig.comparemodel}), as it is naturally expected. Moreover, in every configuration there is a monotonic growth of the velocity anisotropy values with the clustercentric distance, outlining a predominance of radial orbits in the outskirts of the cluster and more isotropic orbits towards the centre. 

\subsubsection{Major populations}

The behaviour of the major cluster members populations (red, blue, high mass, low mass galaxies and all together) is, \textbf{at most radii}, not depending on the population itself, and all their corresponding anisotropy profiles result consistent with each other. We find that the anisotropy values are most sensitive to the choice of the projected number density model, and they are, in general, higher when we adopt the NFW model and lower with the Hernquist model. Among all galaxy populations, the red one is the least sensitive to $N(R)$ changes, while the low mass one is the most influenced. The general sample is also sensitive to $N(R)$ changes.

We note a more marked discrepancy between red and blue galaxies at large radii. Nevertheless, the entity of such discrepancies is highly dependent on the chosen models for $M(r)$, $N(R)$, and $\beta(r)$. Overall, we can conclude that the major galaxy populations follow almost the same anisotropy profile, although there are hints of a possible differentiation between red and blue cluster members.

\subsubsection{Minor populations}

The $\beta(r)$ profiles of minor cluster member populations (red with high $M_*$, red with low $M_*$, blue with high $M_*$, blue with low $M_*$) do not exhibit significant differences in terms of anisotropy values, and we could get to almost the same conclusions we obtain for the major populations. Due to the smaller number of objects in each class, the results for the minor populations are naturally more sensitive to the model choice and they have larger confidence intervals, limiting our possibilities to observe significant differences between them.

\subsection{Jeans inversion results} \label{subsec.Jeansinvres}

As for the \texttt{MAMPOSSt} results, we show in Figure \ref{fig.inversiongeneral} the velocity anisotropy radial profiles resulting from the Jeans inversion, for both of the considered total mass profiles. This technique allows a sharper inspection of the velocity anisotropy profile, due to its unparametrised formulation, overcoming the rigidity of a fixed functional form as those implemented in the \texttt{MG-MAMPOSSt} code. 

The general trend followed by the resulting $\beta(r)$ is similar to that outlined by the \texttt{MG-MAMPOSSt} analysis: we find again more isotropic orbits towards the centre of the cluster, and more radial orbits towards the outskirts of the ensemble cluster. In addition to this, the Jeans inversion results present a very interesting feature at intermediate radii. In every studied population (as well as in the whole sample of cluster members), indeed, we observe a local maximum in $\beta(r)$ at $\sim 250$ kpc from the cluster centre, and a consequent local minimum at $\sim r_{-2}$, while after that point $\beta(r)$ starts increasing again, returning to the original growing trend.
As we find in the \texttt{MAMPOSSt} analysis, there are no significative changes between the velocity anisotropy profiles of different cluster member populations, since the $1\sigma$ confidence intervals of the presented $\beta(r)$ are very often overlapping. However, in concordance to what we show in the previous subsection for \texttt{MAMPOSSt}, these $\beta(r)$ profile might suggest that red galaxies have more radial orbits in the outskirts. Anyway, according to the JI analysis these differences are too small to be conclusive, and we can state that the orbital anisotropy of cluster members does not depend on the chosen galaxy population.

The two tested total mass profiles do not produce significantly different $\beta(r)$ results. Nevertheless, this fact matches our expectations: since the choice of the mass model is just a way to represent the same true total mass profile of a cluster, the outcomes of the two models are not very different.

\section{Discussion} \label{sec.discussion}

So far several studies have investigated the velocity anisotropy profiles of different cluster member populations, both from the observational and simulation point of view.

Starting from the simulation side, much effort has been put in the analysis of the general anisotropy properties of dark matter halos \citep{Wojtak_2005_DManisotropy, Hansen_2006_DManisotropy,Ascasibar_2008_DManisotropy,Lemze_2012_DManisotropy}, while, only most recently, there has been a focus on more specific cluster member populations \citep{Lotz_2019_quenching_Dmanisotropy_simulations,He_2024_anisotropysim,Abdullah_2025_DManisotropy}. The results of these cosmological $N$-body simulations consistently agree on what the typical $\beta(r)$ values in galaxy clusters should be: a growing profile with central values $\beta_0 \gtrsim 0$ and $\beta(R_{200})\approx 0.5-0.7$. As we showed in Section \ref{sec.results}, the anisotropy profiles that we find in the \texttt{MAMPOSSt} analysis provide an excellent confirmation of the simulation results, and the JI results, except for the central peak at $R\sim 0.25$ Mpc, confirm this trend as well. The general growth of the anisotropy values can be observed in almost any configuration that we considered (Figures \ref{fig.MAMmajorgOM}, \ref{fig.MAMmajorgT}, \ref{fig.MAMminorgOM}, \ref{fig.MAMminorgT}), and the values that we find are also very similar to the corresponding mock populations (see, for reference, Figures 11 and 13 in \citealt{Lemze_2012_DManisotropy} or the upper-right panels of Figure 3 in \citealt{Lotz_2019_quenching_Dmanisotropy_simulations}).

On the other hand, the observed anisotropy profiles present a wider phenomenology, probably due to the variety of sample characteristics explored in the literature. 
\cite{Biviano_2004_ENACSanisotropy} studied the ESO Nearby Abell Cluster Survey (ENACS; \citealt{Katgert_1996_ENACSsurvey}) sample by subdividing the cluster members into morphological categories. They determined the total mass profile, using elliptical and S0 kinematics, under the assumption of isotropic orbits, which was supported by the analysis of the shape of their velocity distribution \citep{Katgert_2004_MLratioENACS}. They found that spirals move on radial orbits (see Figures 6 and 8 therein).
It is interesting to compare their results on spiral galaxies with our blue sample and its subdivisions into high- and low-mass blue galaxies: we can see that, although we find anisotropy values slightly higher than theirs for these populations, our blue sample (especially the low mass part, which is the majority) is generally more isotropic than the other populations, at variance with what suggested by \cite{Biviano_2004_ENACSanisotropy}. 
We think that this difference could be ascribed to the different categorisation of cluster member populations, based on morphology rather than on colour. Another possibility is that \cite{Biviano_2004_ENACSanisotropy}'s isotropic assumption for elliptical and S0 orbits is not completely fulfilled. Another possibility is an evolution with redshift of the anisotropy profiles, since our clusters are located at higher $z$ than those of \cite{Biviano_2004_ENACSanisotropy}. However, a more recent study by \cite{Mamon_2019_WINGSanisotropy} did not find a significant difference in the orbits of cluster galaxies of different morphologies at lower redshifts, using the spectroscopic dataset of the clusters of the Wide-field Nearby Galaxy-clusters Survey (WINGS) \citep{Cava_2009_WINGSspeccat,Moretti_2014_WINGScatalog} that is limited to $0.04<z<0.07$. Also in this low-$z$ sample, $\beta(r)$ shows the same growing trend with $r$ for all considered cluster galaxy populations.

An argument in favour of the redshift dependence of the anisotropy profile was presented by
\cite{Biviano_2009_anisotropiesnearfar}   (see Figure 4 therein), where their high-$z$ sample is more similar to ours, since the considered clusters have $0.393 < z < 0.794$ (albeit the considered cluster mass values $M_\mathrm{200c}$ are in the range $0.7 - 13.6\times 10^{14} \, \mathrm{M_\odot}$, with a mean (median) value of $2.8 \,(4.4)\times 10^{14} \, \mathrm{M_\odot}$). They divided their cluster members sample into two galaxy populations (by the presence/absence of emission-lines in their spectra) and they did not find significant differences between the corresponding anisotropy profiles: our extensive work confirms this result, bringing much stronger evidence towards the common shape of $\beta(r)$ regardless of the galaxy population. 

Two other very recent works \citep{Valk_2025_anisotropySDSSlowz,Pizzuti_2025_CHEXMATEanisotropy} investigated the orbital behaviour of different cluster members populations, adopting the same techniques as ours. \cite{Valk_2025_anisotropySDSSlowz} studied a stack of 642 low redshift ($z\lesssim0.2$) galaxy clusters from SDSS, both with \texttt{MAMPOSSt} and JI, which were divided into relaxed and non relaxed clusters according to a gaussianity criterion applied on the observed LOS velocity probability distribution. The relaxed sample, which has an overall $\beta(r)$ profile strongly consistent with other low-$z$ samples such as \cite{Biviano_2009_anisotropiesnearfar} and \cite{Mamon_2019_WINGSanisotropy} (see upper panel of Figure 4 therein), was further divided into 4 populations: star forming, AGNs, transition objects and quiescent. All four populations show very similar anisotropy values (Figure 6), confirming the growing trends and corroborating the evidence of a "universal" anisotropy profile. \cite{Pizzuti_2025_CHEXMATEanisotropy} analysed with \texttt{MG-MAMPOSSt} a stack of 75 clusters from the CHEX-MATE sample \citep{CHEX-MATE_2021_paper}, spanning a wide redshift range up to $z=0.55$. They found a possible hint of a correlation between cluster mass and anisotropy profile.

To sum up, from the comparison of our results to other relevant works in this field, we conclude that there is no significant evidence for different $\beta(r)$ for different cluster galaxy populations, in the explored redshift ranges. In the companion work by \cite{Biviano_2025_CLASHanisotropisingle} (which is based on the same sample used here), we find evidence of $\beta(r)$ evolution of the total cluster member population by comparing the \cite{Biviano_2025_CLASHanisotropisingle} anisotropy profiles with those of galaxy clusters at $z \simeq 0$ \citep{Mamon_2019_WINGSanisotropy, Valk_2025_anisotropySDSSlowz}. Specifically, orbits appear to become more isotropic with time. Instead, it is more difficult to conclude if different galaxy populations evolved their orbits together or at separate times. While \cite{Biviano_2009_anisotropiesnearfar} suggested $\beta(r)$ evolution for the quiescent population and not for the star-forming one, their result is not statistically significant. As a matter of fact, both in our sample and in the higher-$z$ sample of \cite{Biviano_2021_GOGREEN&CLUMPS} the red/quiescent populations have $\beta(r)$ values slightly higher than those of the blue/star-forming populations, at least outside the cluster centre. However, in the low-$z$ sample of \cite{Mamon_2019_WINGSanisotropy} the early-type galaxies display a lower $\beta(r)$ with respect to the late-type galaxies. Comparing these results is not straightforward because of the different characterisation of the different cluster galaxy populations, based on colour, morphology, or star-formation activity, and because the different samples have different stellar mass limits.

\subsection{Orbital evolution and cluster history}

The general framework that is suggested by the comparison between our work and the existing literature may lead to a comprehensive interpretation of a galaxy cluster history based on its orbital behaviour throughout time. Numerical simulations \citep[see, e.g.,][]{Lemze_2012_DManisotropy,Abdullah_2025_DManisotropy} show that the orbital behaviour of cluster members, i.e. $\beta(r)$, depends on the mass of the considered cluster, as well as on its redshift and relaxation status. Particularly, the higher is the cluster mass or redshift, the higher are its anisotropy values. 

The anisotropy-relaxation relation is more complex, and may depend on the phase of the relaxation process. Collisions and major mergers disrupt the phase-space distribution of cluster members, causing violent relaxation \citep{LyndenBell_1967_violentrelax}, that erases ordered motions and causes isotropisation of the orbits \citep{Lemze_2012_DManisotropy}. After a dynamical time, the cluster relaxes and proceeds with a smooth accretion process of field galaxies from surrounding filaments, on radially elongated orbits, so radial anisotropy develops again but mostly in the external cluster regions \citep{Lapi_2011_Relaxation}.
Therefore, it is not trivial to determine the interplay between these processes and which prevails on the others. 

However, the measurements made in the last $\sim20$ years on different - by mass, redshift, and relaxation status - galaxy cluster samples can help to draw a picture of the clusters dynamical history and evolution. At high redshifts, right after the birth of a cluster, its total mass is low, the virialisation has not been completed yet, and the orbits are almost isotropic \citep{Biviano_2016_highzanisotropyGCLASS,Biviano_2021_GOGREEN&CLUMPS}. At intermediate redshifts, as we observe in this work, the virialisation process has been fully completed \footnote{At the mean redshift of the \cite{Biviano_2021_GOGREEN&CLUMPS} sample, $\bar{z}_\mathrm{B}\approx1.1$, the lookback time is $\approx 8.1$ Gyr, while at our mean redshift, $\bar{z}_\mathrm{M}\approx 0.32$, the lookback time is $\approx 3.6$ Gyr.} and a galaxy cluster had enough time to accrete and raise its total mass through cosmic filaments. The newly accreted cluster members are naturally moving along very radial orbits, and this fact makes anisotropy values to raise at all radii. At very low redshifts, there has been enough time to allow the majority of the observed clusters to undergo multiple collisions or major mergers, which can bring them back to a more isotropic state, as it is observed e.g. by \cite{Mamon_2019_WINGSanisotropy}.

\subsection{Methods comparison} \label{subsec.comparison}

The two presented methods offer different points of view on the same problem: the \texttt{MG-MAMPOSSt} code performs Monte Carlo Markov Chain fitting of parametric $\beta(r)$ profiles, inspired by the suggestions coming from cosmological simulations (see e.g. \citealt{He_2024_anisotropysim}), while the non-parametric formulation of the Jeans inversion allow for a more flexible characterisation of the orbital anisotropy. The \texttt{MAMPOSSt} and Jeans inversion results are in agreement for most of the considered projected radii, depending on the choice of the models used in \texttt{MAMPOSSt}, which, in some cases, can create discrepancies localised at some radii (see Figures \ref{fig.MAMmajorgOM}, \ref{fig.MAMmajorgT}, \ref{fig.inversiongeneral}). For convenience, in Figure \ref{fig.comparemodel} we overlap the gT, gOM and Jeans inversion result for the global population as an example. From this plot, we can clearly see that the two methods are in total agreement when characterising the anisotropy profile at large radii, denoting highly radial orbits ($\beta'(r)\approx 2$) in that region. On the other hand, in the innermost regions both methods highlight that orbits are closer to isotropy than in any other region of the cluster. However, between 0.5 and 1.5 Mpc, we find some differences for the two methods: although the confidence intervals of the profiles are still slightly overlapping, the monotonic trend observed in the \texttt{MAMPOSSt} profile is not followed by the Jeans inversion one. The shape of the JI $\beta(r)$ has the peculiarity of a sudden change of direction after a very steep increment of the anisotropy values in the first $\sim 250$ kpc, followed by an intermediate region of more isotropic orbits that is later reconnected to the outer region with higher anisotropy values. The built-in \texttt{MG-MAMPOSSt} $\beta(r)$ profiles are too "rigid" to capture such a feature, and they can catch only the general trend of it. Since the NFW and the Hernquist profiles are robustly verified to be good one-component total mass models for clusters of galaxies (e.g. Appendix C in \citealt{Bonamigo_2018_RXCJ2248_M0416_M1206}; \citealt{Maraboli_2025_virialquantitiesfromSL}), they are as well a suitable input for the Jeans inversion process. This fact could indicate the JI as a more accurate probe for the velocity anisotropy profile, especially in this analysis of a very rich ensemble cluster. To further check if this discrepancy comes from the rigidity of the \texttt{MG-MAMPOSSt} $\beta(r)$ models of Eq. \ref{eq.beta}, we implemented and tested a new ansatz for the anisotropy, which is designed to account for the bump found in the JI results (Pizzuti-Biviano, hereafter BP, \citealt{Pizzuti_2025_BPbetamodel}):
\begin{equation}
    \label{eq.BPmodel}
    \beta_\mathrm{BP}(r)=\beta_0+(\beta_\infty - \beta_0)\left[\frac{r}{r+r_\beta}+\frac{r^2}{r^2_\beta}e^{-\left(\frac{r}{r_\beta}\right)^2}\right],
\end{equation}
where $r_\beta$ defines the position of the bump. We considered $r_\beta$ as a free parameter in the \texttt{MG-MAMPOSSt} fit. We show the results in Figure \ref{fig.comparemodel} for the total population. We checked for the significance of the JI feature by comparing the \texttt{MG-MAMPOSSt} BP and gT model results with the Bayes Information Criterion (BIC; \citealt{Schwartz_1978_BIC}). With both the NFW and the Hernquist total mass profiles, we find $\Delta\mathrm{BIC}=\mathrm{BIC_{BP}}-\mathrm{BIC_{gT}}\sim6$, indicating that the gT model is strongly favoured compared to the BP model \citep{Kass_1995_BICsignificance}, which makes us question the physical reality of the $\beta(r)$ feature found in the JI analysis.

Moreover, we carefully investigated the possible origin of the JI feature with additional tests: 1) we rescaled with $r_{-2}$ instead of $R_\mathrm{200c}$, but this parametrisation resulted to be weaker since $r_{-2}$ is more poorly constrained than $R_\mathrm{200c}$ from the data of \cite{Umetsu_2018_CLASHmassWL} (mostly because of the large errors on the values of concentration); 2) we tested if one of the clusters composing the stack was "problematic", and we reran the JI analysis removing one cluster at a time from the stack; 3) we tested if the JI feature was an artifact of the interpolation to $R=0$ made by the JI analysis software, by recomputing the velocity dispersion profile at smaller radii to limit the interpolated radial range. However, none of these tests succeeded in removing the JI $\beta(r)$ feature.

We conclude that while the data, taken at face value, indicate the existence of the $\beta(r)$ feature, the statistics provided by our dataset does not allow for a robust confirmation of it.

\subsection{General interpretation}

The behaviour of the velocity anisotropy profiles that we find with both the adopted techniques can be brought back to some known physical processes that involve galaxy clusters.

Starting from the outer regions of the ensemble cluster, the highly radial orbits that we find in that area can be traced to the new infalling galaxies that recently joined the cluster. This consideration explains why the velocity anisotropy values at those radii are higher than in any other region of the cluster: when a galaxy falls from the surrounding environment towards the cluster centre, its orbit is mostly radial; instead, when we consider regions that are below the virial radius, cluster members begin to be more and more isotropic.

In the innermost regions of the cluster, below $\sim 250$ kpc, we find velocity anisotropy values that are the lowest with respect to the others. This time, we reconnect this specific behaviour to the presence of dynamical friction \citep{Chandrasekhar_1943_dynamicalfriction}, because at these radii the cluster environment is very crowded (as we can see from Figure \ref{fig.Nprofiles}). Since the dynamical friction force is directly proportional to the velocity of the object experiencing it, the principal component of the orbital velocity will be affected the most by friction, whether it is the tangential or the radial component. Hence, this process leads the orbital velocity components to be roughly equal, and so it leads to isotropy.

At intermediate radii, where the two methods produce different results for $\beta(r)$, the \texttt{MAMPOSSt} outcomes naturally return a smooth transition between the radial orbits of the cluster outskirts and the more isotropic ones at the centre. The JI results, instead, offer a more intriguing view on the dynamics of cluster members at intermediate radii, between 250 and 1500 kpc. 

\subsection{An insight on the methodologies}

The techniques that we adopt to study the velocity anisotropy profiles have been thoroughly tested and have been proven to be robust estimators of the $\beta(r)$ profiles (e.g. \citealt{Biviano_2004_ENACSanisotropy,Biviano_2013_M1206,Biviano_2024_jellyfishorbits,Annunziatella_2016_A209speccat,Mamon_2019_WINGSanisotropy,Sartoris_2020_AS1063,Biviano_2025_CLASHanisotropisingle}). As every method, there are some aspects that we believe are worth examining to better comprehend the results that we have presented in Section \ref{sec.results}. Further discussion about the Jeans inversion technique can be found in Appendix \ref{app.comparedissect}.

First, we want to address how the adoption of the mass profiles described in Section \ref{subsec.masscraft} affects our \texttt{MAMPOSSt} analysis, since the scale radii (and their confidence intervals) determined in that Section are directly involved in the priors given to \texttt{MG-MAMPOSSt}. The code, as we report in Section \ref{subsec.mamposst}, fixes the scale radius of the selected beta model $r_\beta$ to the $r_{-2}$ of the total mass profile. We set the prior for the total mass scale radius $r_\mathrm{s}$ as a uniform distribution between $r_\mathrm{s}^\mathrm{guess}-\sigma_r$ and $r_\mathrm{s}^\mathrm{guess}+2\sigma_r$, where $r_\mathrm{s}^\mathrm{guess}$ is set as the value of $r_\mathrm{NFW}$ or $r_\mathrm{H}$ (depending on the chosen configuration) computed in Section \ref{subsec.masscraft}, and $\sigma_r$ is its corresponding error.

\subsubsection{Jeans inversion error contributions}\label{subsec.jeanserrs}
Another interesting topic about the total mass profile for the stacked cluster is how this choice impacts the systematic errors of our results. The confidence intervals that we report in Figure \ref{fig.inversiongeneral} are, indeed, the final product of many factors of uncertainty. We identify and characterise two main sources of it:
\begin{itemize}
    \item The set of arbitrarily chosen parameters that were to be made in order to compute the $\beta(r)$ profile, as they represent the most relevant component of the systematic errors in our analysis. This set includes the smoothing factors for $N(R)$ and $\sigma_\mathrm{los}(R)$, the constants $\xi$ and $f$ employed in Equation \ref{eq.Nextrap} and \ref{eq.sigmaextrap}, respectively, the infinity radius $R_\mathrm{inf}$, the scale radii of the total mass profiles $r_\mathrm{NFW}$ and $r_\mathrm{H}$.
    \item The statistical errors that naturally come from our data, namely from the velocity dispersion radial profile, $\sigma_\mathrm{los}(R)$, and the projected number density profile, $N(R)$.
\end{itemize}
Although the uncertainties related to the total mass profiles that we adopt for the Jeans inversion are related to those in \cite{Umetsu_2018_CLASHmassWL}, we include them among the systematics because of the profile building process that we describe in Section \ref{subsec.masscraft}. The resulting total mass profile, indeed, is determined based on a set of reasonable, but arbitrary, choices about its fundamental quantities, such as the definition of the ensemble cluster concentration $\Bar{c}_\mathrm{200c}$ (Equation \ref{eq.concentration}) and the definition of its $r_{-2}$.

In Figure \ref{fig.jeanserr} we report, as an example, the velocity anisotropy profile for the whole sample of blue cluster members, with the confidence intervals dissected according to the characterisation explained above. From this plot, it is clear that the bootstrap of the $N(R)$ and $\sigma_\mathrm{los}(R)$ profiles is the main source of uncertainty, while the systematic one is marginal. We also notice some correlation between the two kinds of errors, since the general confidence interval width does not correspond to the sum of the two interval widths.

\subsubsection{Variance of the cluster sample}

Our cluster sample offers the rare opportunity to be analysed both through the stacking of its components, as we do in this work, and in a cluster-by-cluster way by \cite{Biviano_2025_CLASHanisotropisingle}. In the latter there emerges a significant variance in $\beta(r)$ among the considered clusters; however, their average $\beta(r)$ is in good agreement with the average $\beta(r)$ from the \texttt{GAEA} simulated halos of similar mass and redshift, which is also consistent with our results. \cite{Biviano_2025_CLASHanisotropisingle} suggest that this variance could be due to the different cluster orientations relative to their major axis, and we refer to Section 5 of that work for an extended discussion on the variance origins.

\section{Summary and conclusions} \label{sec.sumconcl}

In this paper we compared two robust techniques for measuring the orbital velocity anisotropy profile $\beta(r)$ in a stack of galaxy clusters, namely the \texttt{MAMPOSSt} method and the Jeans inversion. Thanks to our very large number of cluster members, from these two parallel analyses we could obtain a clear, general picture of how the $\beta(r)$ profiles behave for different cluster members populations. We summarise our work here:
\begin{itemize}
    \item We consider a sample of nine massive galaxy clusters at intermediate redshifts ($M_\mathrm{200c}>7\times 10^{14} \, \mathrm{M_\odot}$, $0.18<z<0.45$, see Table \ref{tab.clusters}), selected from the CLASH-VLT \citep{Rosati_2014_CLASH} targeted clusters. Our main target is to study the behaviour of the velocity anisotropy profile $\beta(r)$ for different galaxy populations, and we do it by stacking the cluster members in a renormalised projected phase space (PPS) and considering them as a part of a single ensemble cluster.
    \item The selection of cluster members is based on a recalibrated version of the CLUMPS algorithm, that operates on the galaxy distribution in the PPS and looks for local minima of the surface number density of spectroscopic objects (see Figure \ref{fig.membership}). The selected members are then labelled as red or blue galaxies depending on their belonging to the red sequence (Section \ref{subsec.colorsub}). They are also labelled as high/low stellar mass galaxies according to their I-band magnitude (Section \ref{subsec.masssub}).
    \item We stack the member galaxies of different clusters in the same PPS by normalising their distance from the cluster centre by the corresponding cluster $R_\mathrm{200c}$, and their rest-frame velocity by $v_\mathrm{200c}=\sqrt{GM_\mathrm{200c}/R_\mathrm{200c}}$. Then, we rescale the galaxy PPS coordinates by $\langle R_\mathrm{200c} \rangle=2.18$ Mpc ($\langle \cdot \rangle$ stands for average weighted with respect to the number of cluster members) and $\langle v_\mathrm{200c} \rangle=1732$ km/s, respectively.
    \item After the stacking procedure, we produce the surface number density profiles $N(R)$ (Figure \ref{fig.Nprofiles}) for the different cluster member populations. This required a proper accounting of the spectroscopic completeness of our sample (Section \ref{subsec.speccomp}), that is consequently employed to weight the contribution of each cluster member to $N(R)$.
    \item Finally, we analyse the cluster members employing two different methods: the \texttt{MAMPOSSt} method (Section \ref{subsec.mamposst}), that performs a MCMC fit of the PPS adopting the parametric profiles for $N(R)$, $M(r)$ and $\beta(r)$ presented in Equations \ref{eq.NFW}, \ref{eq.hernq}, \ref{eq.nfwmass}, \ref{eq.hernmass}, and \ref{eq.betageneral}; the Jeans inversion (Section \ref{subsec.jeansinv}), that returns a non-parametric measure of $\beta(r)$ upon providing the total mass profile from Equations \ref{eq.nfwmass} and \ref{eq.hernmass} as an input. 
\end{itemize}

The results of these two parallel analyses outline a growth of the anisotropy values from the centre to the outskirts for every considered cluster member population, with $\beta'=\sigma_r/\sigma_\vartheta$ values of $\lesssim 1.4$ at the very centre and of $\gtrsim 2$ at the virial radius. These values can vary depending on the adopted models and methods, but the general trend is always confirmed. We find a possibly interesting feature in all the $\beta$ profiles from the Jeans inversion technique, that is (see Figure \ref{fig.inversiongeneral}) the sudden switch at $250-400$ kpc from rapidly increasing anisotropy values to a mild, but evident, decrement in the following $500-700$ kpc, and ending up with a final increase that ends up matching the high anisotropy values found in the \texttt{MAMPOSSt} analysis.

We do not find significant differences in $\beta(r)$ among the studied galaxy populations, although some differences are suggested for certain choices of $M(r)$, $N(R)$, and $\beta(r)$ between the red and blue populations at large radii. As we discussed in Section \ref{sec.discussion}, this behaviour is consistently maintained throughout the evolution of a galaxy cluster, which may change the overall orbital anisotropy of its components.

The very high precision of our results, based on an unprecedented number of cluster member redshifts, and the robustness of our techniques make the obtained $\beta(r)$ profiles a reliable measurement of velocity anisotropy in clusters of galaxies. The mathematical simplicity of the anisotropy profiles found with the \texttt{MAMPOSSt} method makes them a suitable and "ready-to-use" tool for future dynamical mass measurements. On the other hand, the freedom allowed by the Jeans inversion in determining the true shape of the velocity anisotropy profile sheds light on potential new interesting dynamical features of galaxy clusters. Measuring such features could improve our current knowledge of galaxy cluster structure and, consequently, provide new inputs for diverse cosmological studies that rely on them.

\bibliography{bibliografia.bib}

\onecolumn

\begin{appendix}
\section{Colour cuts for completeness study} \label{app.colorcuts}

Since we want to evaluate the radial completeness of the cluster members, we introduce the following two colour cuts, chosen by visually inspecting the position of the spectroscopic cluster members, selected before, in two different colour diagrams:
\begin{itemize}
    \item [1)] The first cut is applied in the colour-colour plane defined by B$-$R on the vertical axis and V$-$I on the horizontal one. The cluster region is selected to maximise the purity of the sample enclosed within. We distinguish two main regions onto which we apply the cut, corresponding approximately to the population of red cluster galaxies and to the blue/green one, respectively. In Table \ref{tab.cut1} we collect all the details about the cuts on the colour-colour plot of each galaxy cluster, and every cut is made as follows: we choose the galaxies with $A_k(\mathrm{V-I})+B_k \leq \mathrm{B-R} \leq A_k(\mathrm{V-I})+B_k +$Offset, where $A_k$, $B_k$, and "Offset" are different coefficients for each region, and where V$-$I spans in the intervals specified in Table \ref{tab.cut1}. The purity of the selection is then computed as the ratio between the number of cluster members and the total number of spectroscopic objects in the region.
    \item [2)] The second cut is applied on the subsample of galaxies selected from the first cut, and it is done by looking at the R magnitude vs B$-$R colour plot. Here the main target of the cut is to eliminate carefully the brightest non-member galaxies, that clearly are at lower redshifts than the cluster members. As we did for the first cut, we collect in Table \ref{tab.cut2} the details of the final cut made on each galaxy cluster, in which we choose the galaxies that follow the selection rule B$-$R$\geq \min\{A_\mathrm{final}\mathrm{R}+B_\mathrm{final},\ \mathrm{B-R} \,\ \mathrm{threshold}\}$. We also report the new purities, recomputed after this second cut.
\end{itemize}

\begin{figure*}
\centering
\includegraphics[scale=0.38]{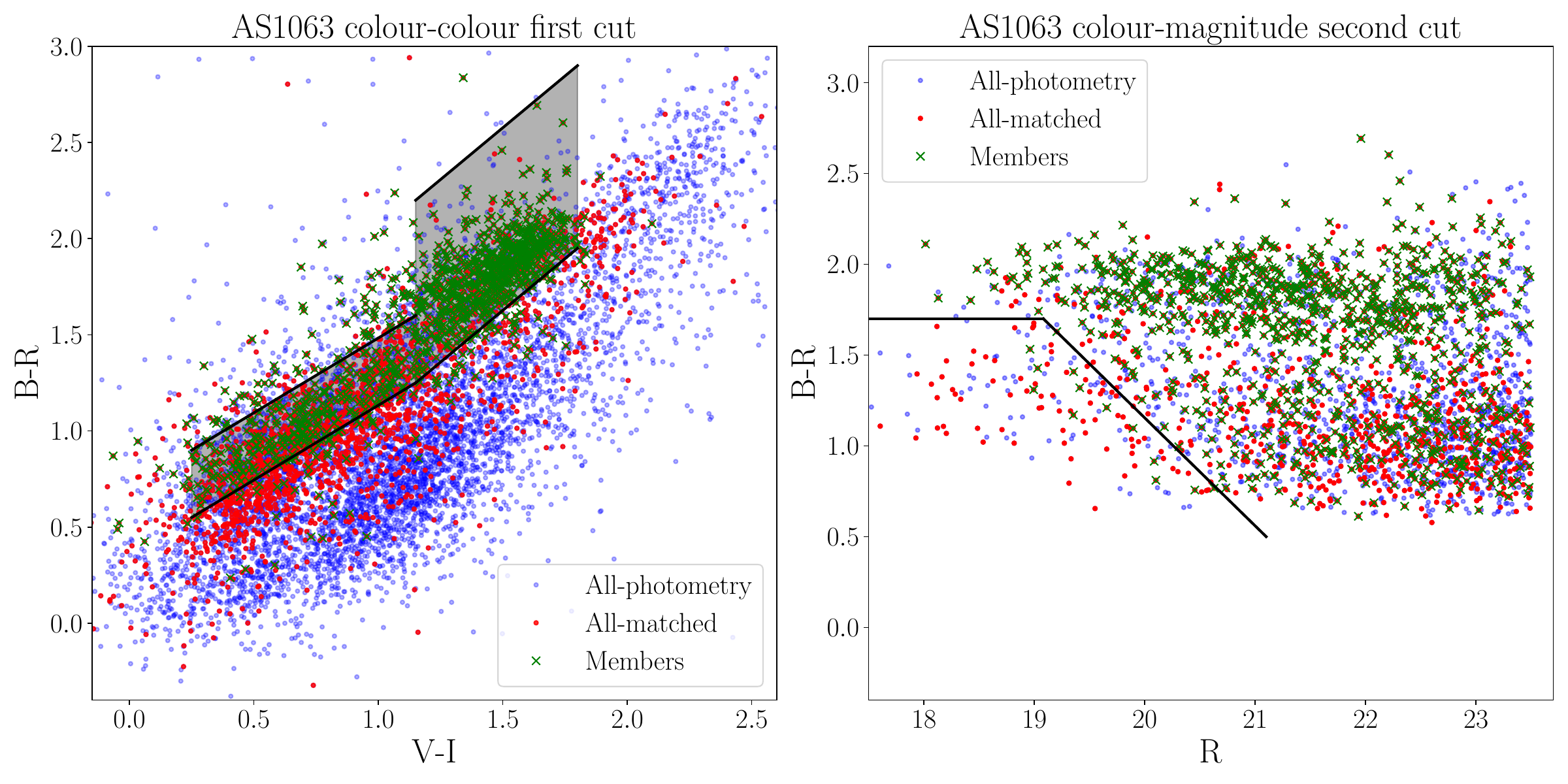}
\vspace{-0.3cm}
\caption{Colour cuts operated on the AS1063 catalogues. In the left panel, we report the first cut (grey region marked by black boundaries) in the colour-colour plane V$-$I versus B$-$R. In the right panel, we report the cut (second cut), operated in the R band magnitude versus B$-$R colour plane, of the subsample extracted from the first cut. The selected objects in this panel are those above the black line. The "All-matched" entry in the legend, corresponding to the red points, indicates the objects in the photometric catalogue that have a match in the spectroscopic catalogue.}
\label{fig.colorcuts}
\vspace{-0.3cm}
\end{figure*}

We report in Figure \ref{fig.colorcuts} an example of the two colour cuts that we apply on AS1063: in the left panel we depict the first cut on the colour-colour plane, while in the right panel we show the second cut in the colour-magnitude plane. The details of every cut can be found in Table \ref{tab.cut1}, in which we also report the purities of the selected subsamples. We remark that although we find an average purity of $\sim 50\%$, it is almost impossible to achieve better values because of the lack of spectroscopic coverage on the outskirts of the considered cluster. Since in these regions the interloper fraction is higher, but also the fraction of blue cluster members, the purity of the blue cluster member population is expected to be lower than that of the red population. Hence, in the blue part of the colour-colour plot we have far less cluster members among the spectroscopic objects, and this heavily affects our average purity. On the other hand, in the red part of the colour-colour plot the purities can easily reach values of $\sim 80\%$, since in the inner regions of the considered clusters (mostly populated by red galaxies) we have a higher chance that a spectrograph pointing catches the majority of cluster members. This purity issue is the reason why we do the second colour-magnitude cut, that allows us to raise the purity by some percentage points, as we show in Table \ref{tab.cut2}.

\begin{table*}[h!]
\caption{Values of the parameters employed for the first (colour-colour) cut.}
\label{tab.cut1}
\centering
\begin{tabular}{cccccccccccc}
\toprule
\multirow{2}{*}{Cluster}  & \multicolumn{5}{c}{Blue region} & \multicolumn{5}{c}{Red region} & \multirow{2}{*}{Overall Purity} \\
\cmidrule(lr){2-6} \cmidrule(lr){7-11} & $V-I$ interval & $A_\mathrm{blue}$ & $B_\mathrm{blue}$ & Offset & Purity & $V-I$ interval & $A_\mathrm{red}$ & $B_\mathrm{red}$ & Offset & Purity \\
\midrule
A383 & [0.28, 0.55] & 0.93 & 0.49 & 0.2 & 27\% & [0.55, 0.93] & 2.5 & $-$0.375 & 0.3 & 36\% & 33\% \\
A209 & [0.3, 0.6] & 1.67 & 0.10 & 0.2 & 38\% & [0.6, 0.95] & 1.43 & 0.24 & 0.7 & 80\% & 72\% \\
R2129 & [0.45, 0.75] & 1.33 & 0.15 & 0.2 & 17\% & [0.75, 1.15] & 1.88 & $-$0.16 & 0.25 & 62\% & 45\% \\
MS2137 & [0.5, 1.1] & 1.08 & 0.21 & 0.2 & 29\% & [1.1, 1.55] & 1.27 & 0.01 & 0.25 & 71\% & 46\% \\
AS1063 & [0.25, 1.15] & 0.78 & 0.36 & 0.35 & 34\% & [1.15, 1.8] & 1.08 & 0.01 & 0.95 & 81\% & 57\% \\
M1115 & [0.3, 1.0] & 0.86 & 0.64 & 0.2 & 31\% & [1.0, 2.1] & 1.33 & 0.17 & 0.5 & 78\% & 51\% \\
M1931 & [0.4, 1.15] & 0.93 & 0.73 & 0.15 & 26\% & [1.15, 1.57] & 0.95 & 0.80 & 0.3 & 75\% & 50\% \\
M1206 & [0.8, 1.55] & 0.93 & 0.00 & 0.25 & 42\% & [1.55, 2.05] & 1.30 & $-$0.52 & 0.45 & 84\% & 66\% \\
M0329 & [0.35, 1.3] & 0.79 & 0.47 & 0.15 & 38\% & [1.3, 1.7] & 1.00 & 0.40 & 0.5 & 71\% & 53\% \\

\bottomrule
\end{tabular}
\tablefoot{ We list the intervals, the inclinations and the resulting purities of the selection for both the red and the blue region of the colour-colour plane (see text). In the last column, we report the combined purity computed as the ratio between the number of cluster members and the total number of spectroscopic objects in the region.}
\end{table*}

\vspace{-0.4cm}

\begin{table}[h!]
\caption{Values of the parameters employed for the second cut.}
\label{tab.cut2}
\centering
\begin{tabular}{ccccc}
\toprule
Cluster & $A_\mathrm{final}$ & $B_\mathrm{final}$ & B$-$R threshold & Final Purity \\
\midrule
A383 & $-$0.44 & 10.2 & 1.7 & 34\% \\
A209 & $-$0.24 & 6.0 & 1.6 & 73\% \\
R2129 & $-$0.40 & 9.5 & 1.85 & 53\% \\
MS2137 & $-$0.44 & 10.2 & 1.8 & 52\% \\
AS1063 & $-$0.59 & 13.3 & 1.7 & 60\% \\
M1115 & $-$0.49 & 11.3 & 1.8 & 56\% \\
M1931 & $-$0.59 & 13.6 & 2.0 & 56\% \\
M1206 & $-$0.88 & 19.4 & 1.7 & 70\% \\
M0329 & $-$0.44 & 10.4 & 1.7 & 60\% \\

\bottomrule
\end{tabular}
\tablefoot{In the last column, we report the overall purity, that improves from the previous cut (see Table \ref{tab.cut1}) for all the considered clusters.}
\vspace{-0.9cm}
\end{table}

\section{Supplementary plots and tables of the MAMPOSSt analysis} \label{app.betaplots}
\begin{figure*}[h!]
\centering
\includegraphics[scale=0.34]{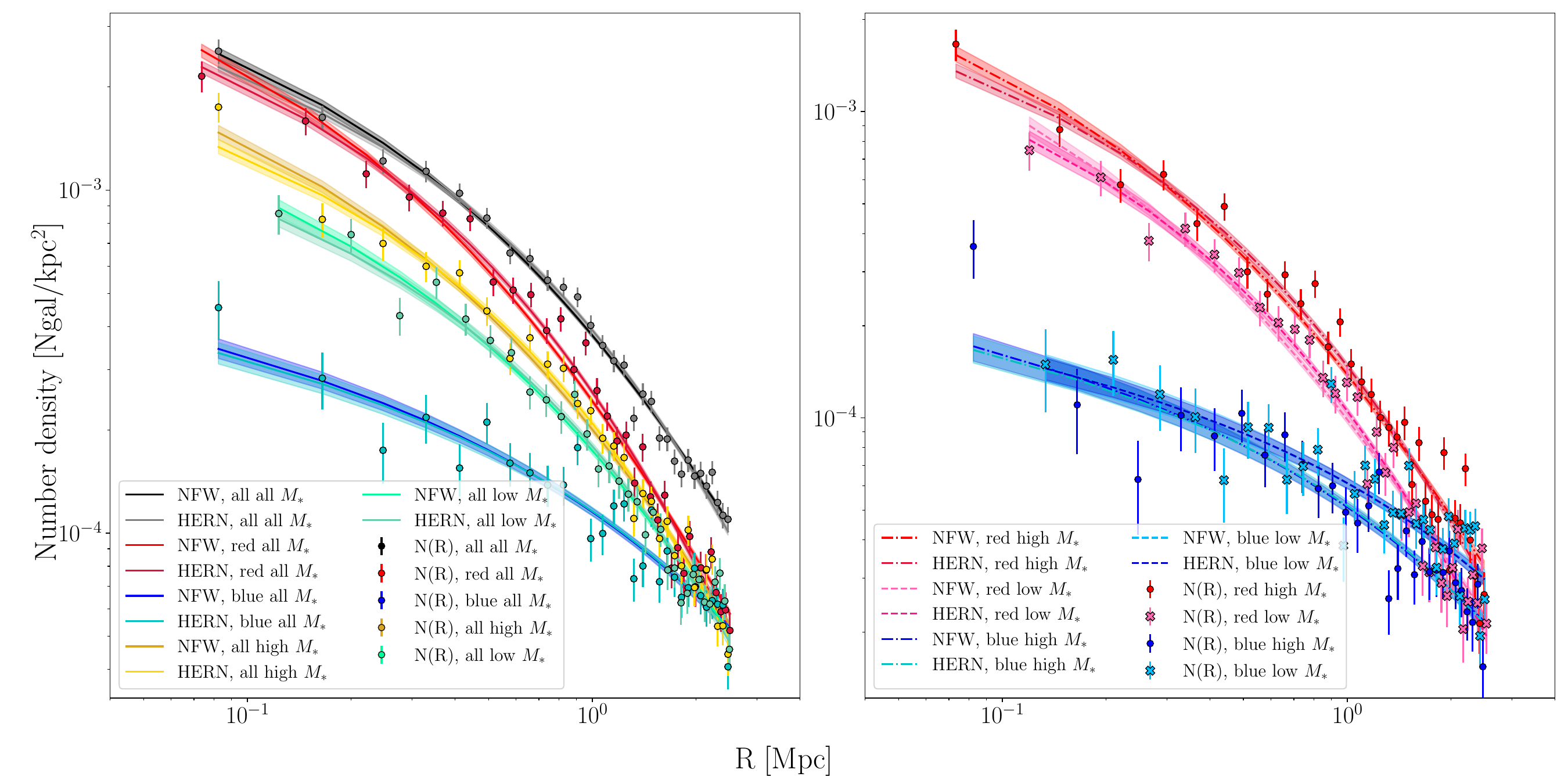}
\vspace{-0.2cm}
\caption{Projected number density profiles $N(R)$ for all the considered subsamples of galaxy populations after stacking. In the left panel, we show the $N(R)$ of the main galaxy populations (all red galaxies, all blue galaxies etc.), as well as the $N(R)$ of the whole sample of selected cluster members. In the right panel, we show the four subcategories of the main populations (red with high $M_*$, blue with low $M_*$ etc.)}
\label{fig.Nprofiles}
\vspace{-0.3cm}
\end{figure*}

\begin{figure*}
\centering
\includegraphics[scale=0.31]{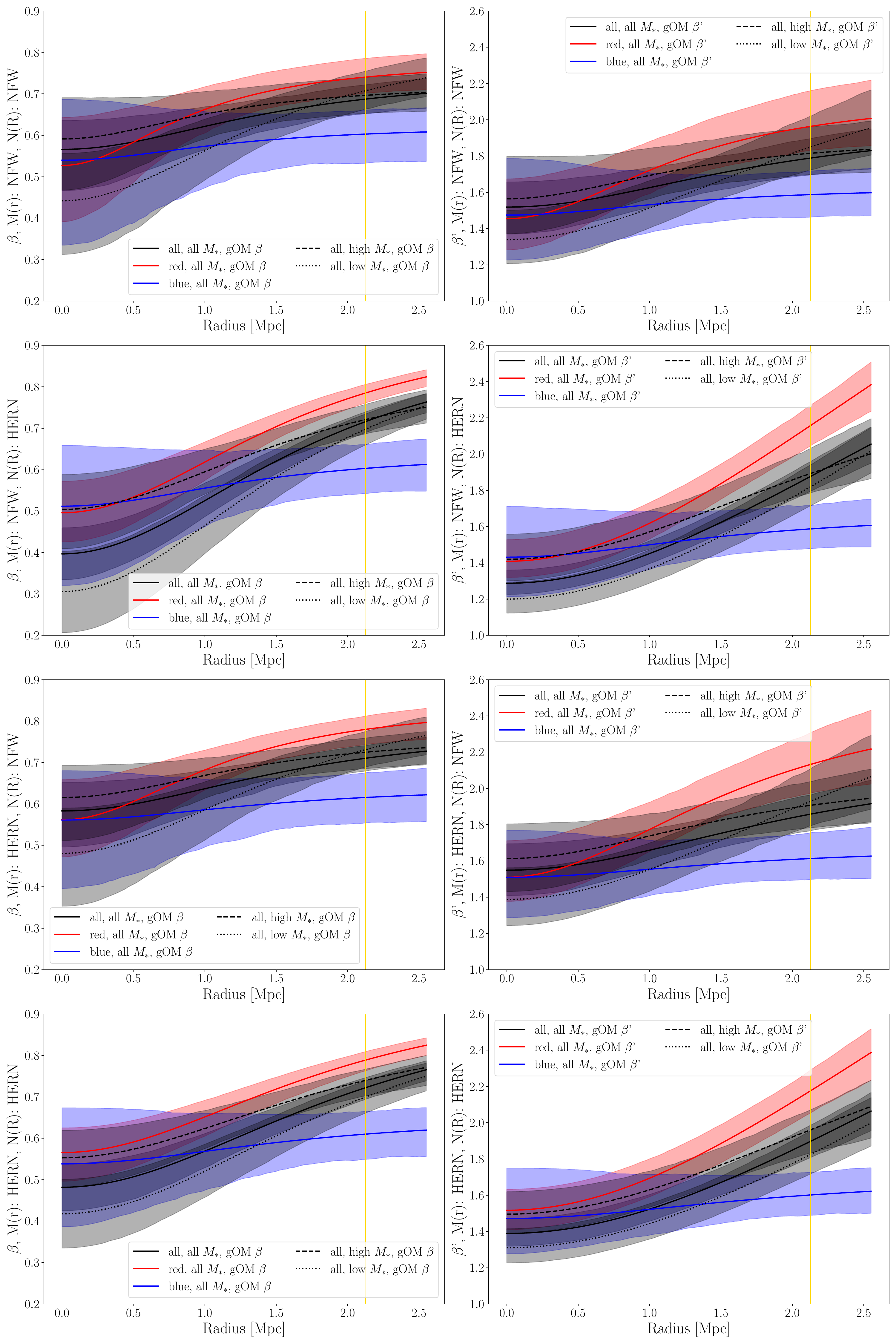}
\caption{Results of the generalised Osipkov-Merritt anisotropy model for the major cluster member populations. In the left column, we show the $\beta(r)$ values, while, on the right one, we plot those of $\beta'(r)=\sigma_r/\sigma_\vartheta$. The four rows are organised to distinguish the different mass and number density models adopted in each situation, and they are specified on the label of each vertical axis. The shadowed region represents the $1\sigma$ confidence interval, while the yellow line represents the $R_\mathrm{200c}$ of the stacked cluster.}
\label{fig.MAMmajorgOM}
\end{figure*}

\begin{figure*}
\centering
\includegraphics[scale=0.31]{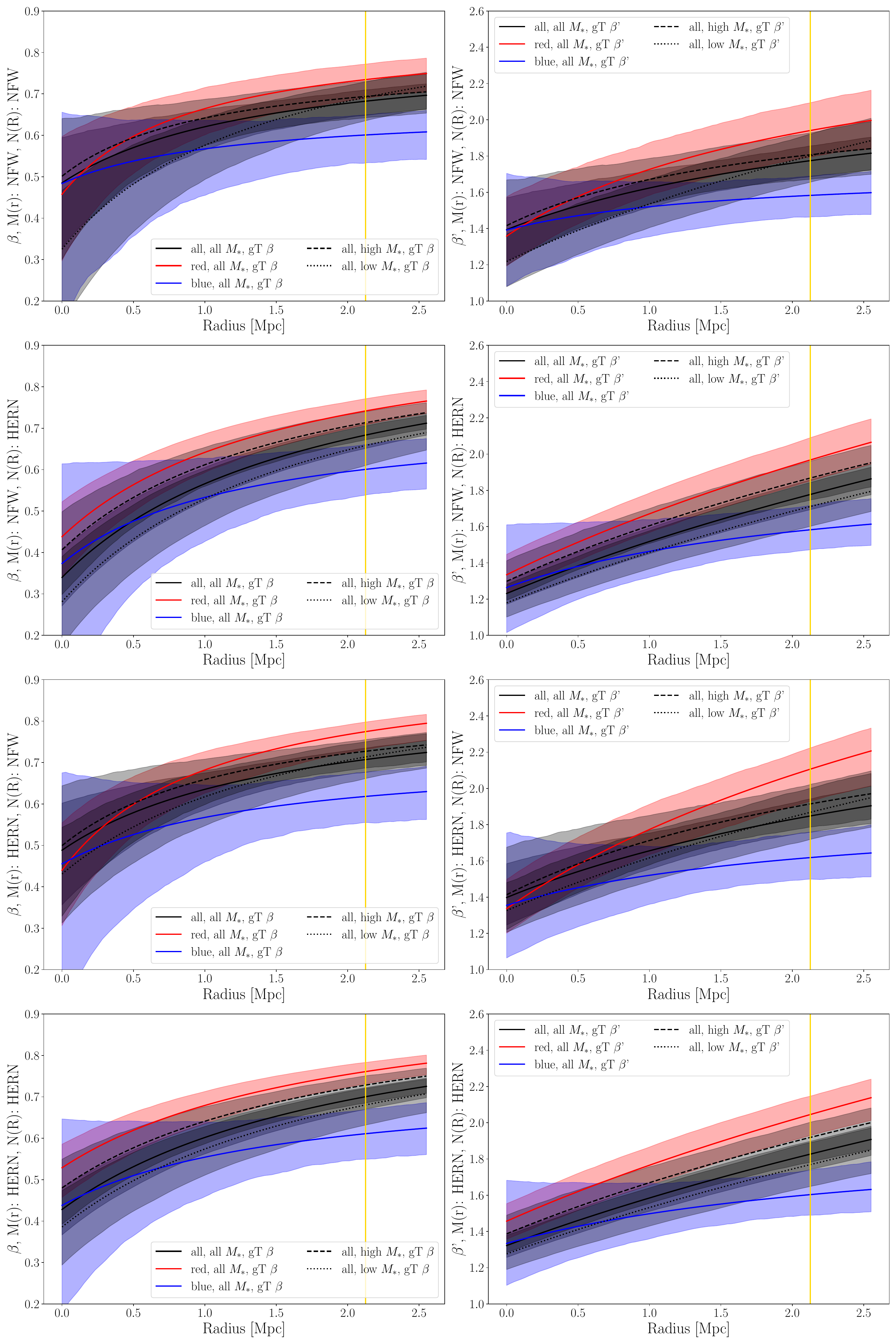}
\caption{Results of the generalised Tiret anisotropy model for the major cluster member populations. See Figure \ref{fig.MAMmajorgOM} caption for further details about this Figure.}
\label{fig.MAMmajorgT}
\end{figure*}

\begin{figure*}
\centering
\includegraphics[scale=0.31]{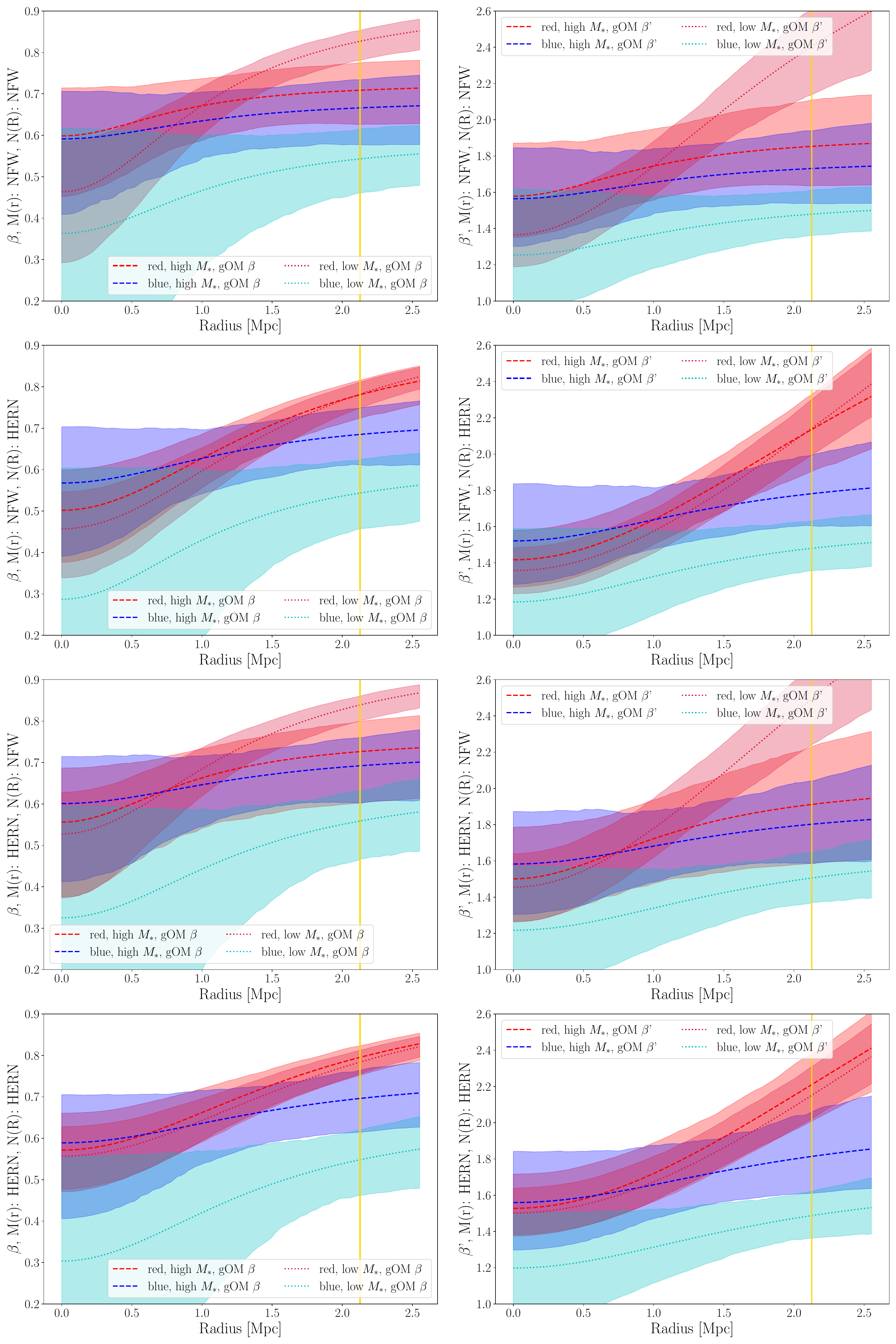}
\caption{Results of the generalised Osipkov-Merritt anisotropy model for the minor cluster member populations. See Figure \ref{fig.MAMmajorgOM} caption for further analogous details about this Figure.}
\label{fig.MAMminorgOM}
\end{figure*}

\begin{figure*}
\centering
\includegraphics[scale=0.31]{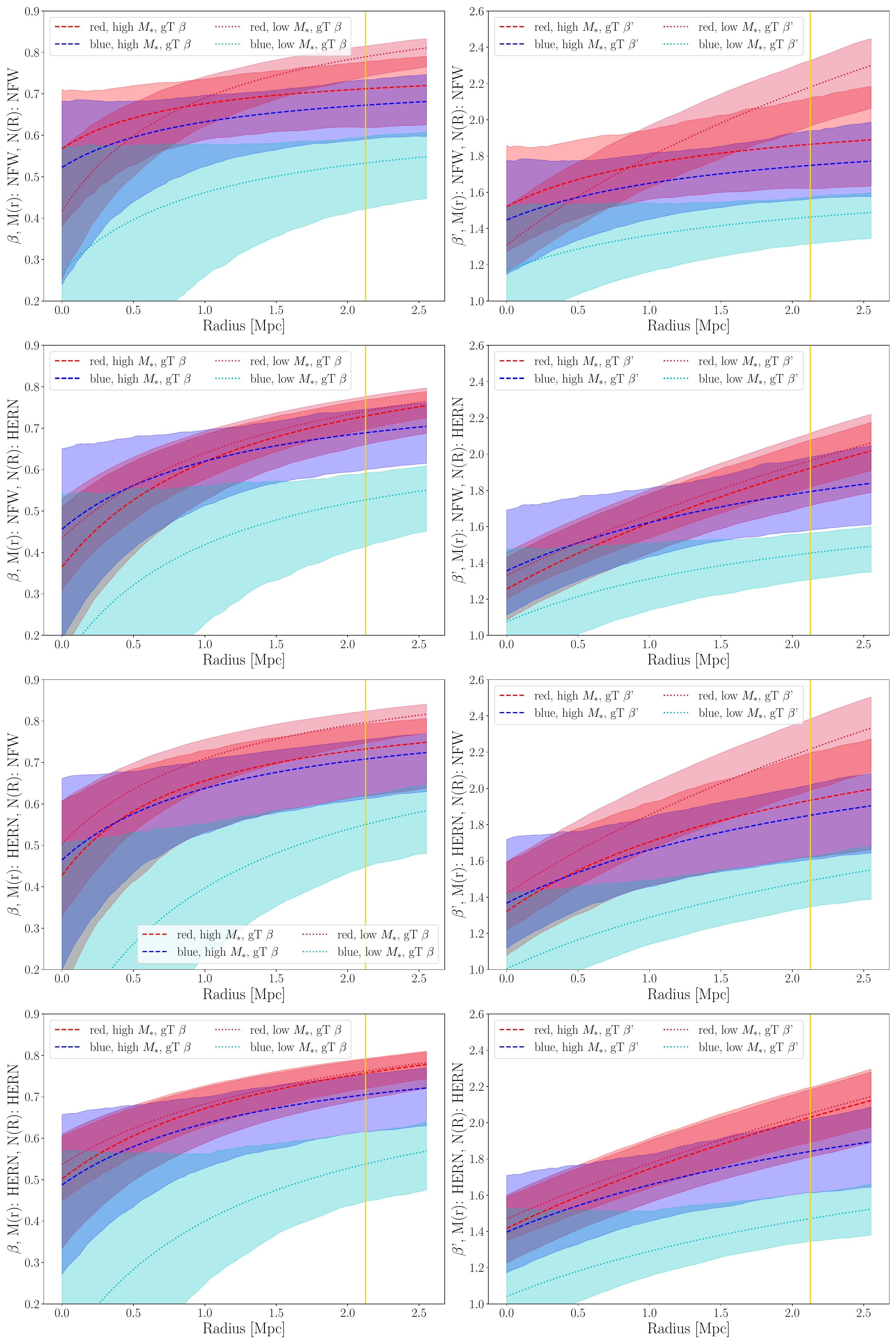}
\caption{Results of the generalised Tiret anisotropy model for the minor cluster member populations. See Figure \ref{fig.MAMmajorgOM} caption for further details about this Figure.}
\label{fig.MAMminorgT}
\end{figure*}

Here, we report supplementary plots and tables of the \texttt{MAMPOSSt} analysis. In Figure \ref{fig.Nprofiles}, we plot the fitted $N(R)$ profiles employed by \texttt{MG-MAMPOSSt}. In Figures \ref{fig.MAMmajorgOM} and \ref{fig.MAMmajorgT}, we show the results for, respectively, the gOM and gT anisotropy model of the major galaxy populations, while in Figures \ref{fig.MAMminorgOM} and \ref{fig.MAMminorgT}, we show the corresponding minor population plots. We also report in Tables \ref{tab.betamajor} and \ref{tab.betaminor} the median values of $\beta_0$ and $\beta_\infty$ with the corresponding $1\sigma$ confidence intervals.
\vspace{-0.1cm}

\begin{table*}
\centering
\caption{Best-fit values of $\beta_0$ and $\beta_\infty$ for the major galaxy populations.}
\vspace{-0.2cm}
\begin{tabular}{cccccccccccc}
\toprule
\midrule
\multicolumn{2}{c}{\textbf{gT}} & \multicolumn{2}{c}{All} & \multicolumn{2}{c}{Red} & \multicolumn{2}{c}{Blue} & \multicolumn{2}{c}{High $M_*$} & \multicolumn{2}{c}{Low $M_*$} \\
\cmidrule(lr){1-2} \cmidrule(lr){3-4} \cmidrule(lr){5-6} \cmidrule(lr){7-8} \cmidrule(lr){9-10} \cmidrule(lr){11-12} $M(r)$ & $N(R)$ & $\beta_0$ & $\beta_\infty$ & $\beta_0$ & $\beta_\infty$ & $\beta_0$ & $\beta_\infty$ & $\beta_0$ & $\beta_\infty$ & $\beta_0$ & $\beta_\infty$ \\
\midrule
NFW & NFW & $0.49^{+0.12}_{-0.15}$ & $0.82^{+0.07}_{-0.08}$ & $0.46^{+0.14}_{-0.16}$ & $0.85^{+0.07}_{-0.08}$ & $0.48^{+0.18}_{-0.29}$ & $0.67^{+0.12}_{-0.13}$ & $0.50^{+0.15}_{-0.19}$ & $0.79^{+0.08}_{-0.09}$ & $0.32^{+0.15}_{-0.17}$ & $0.95^{+0.03}_{-0.07}$ \\
NFW & HERN & $0.34^{+0.06}_{-0.07}$ & $0.98^{+0.01}_{-0.02}$ & $0.44^{+0.08}_{-0.10}$ & $0.98^{+0.01}_{-0.03}$ & $0.37^{+0.24}_{-0.36}$ & $0.74^{+0.15}_{-0.14}$ & $0.41^{+0.09}_{-0.10}$ & $0.95^{+0.03}_{-0.07}$ & $0.28^{+0.10}_{-0.11}$ & $0.98^{+0.01}_{-0.02}$ \\
HERN & NFW & $0.49^{+0.11}_{-0.12}$ & $0.88^{+0.08}_{-0.09}$ & $0.44^{+0.12}_{-0.14}$ & $0.95^{+0.03}_{-0.06}$ & $0.46^{+0.20}_{-0.32}$ & $0.73^{+0.17}_{-0.17}$ & $0.50^{+0.14}_{-0.16}$ & $0.87^{+0.09}_{-0.10}$ & $0.43^{+0.12}_{-0.12}$ & $0.95^{+0.03}_{-0.07}$ \\
HERN & HERN & $0.43^{+0.06}_{-0.06}$ & $0.98^{+0.01}_{-0.03}$ & $0.53^{+0.06}_{-0.07}$ & $0.98^{+0.01}_{-0.02}$ & $0.44^{+0.20}_{-0.28}$ & $0.74^{+0.18}_{-0.17}$ & $0.48^{+0.08}_{-0.08}$ & $0.96^{+0.02}_{-0.07}$ & $0.39^{+0.09}_{-0.09}$ & $0.98^{+0.01}_{-0.02}$ \\
\midrule
\midrule
\multicolumn{2}{c}{\textbf{gOM}} & \multicolumn{2}{c}{All} & \multicolumn{2}{c}{Red} & \multicolumn{2}{c}{Blue} & \multicolumn{2}{c}{High $M_*$} & \multicolumn{2}{c}{Low $M_*$} \\
\cmidrule(lr){1-2} \cmidrule(lr){3-4} \cmidrule(lr){5-6} \cmidrule(lr){7-8} \cmidrule(lr){9-10} \cmidrule(lr){11-12} $M(r)$ & $N(R)$ & $\beta_0$ & $\beta_\infty$ & $\beta_0$ & $\beta_\infty$ & $\beta_0$ & $\beta_\infty$ & $\beta_0$ & $\beta_\infty$ & $\beta_0$ & $\beta_\infty$ \\
\midrule
NFW & NFW & $0.57^{+0.08}_{-0.09}$ & $0.75^{+0.04}_{-0.05}$ & $0.53^{+0.11}_{-0.13}$ & $0.78^{+0.05}_{-0.06}$ & $0.54^{+0.14}_{-0.20}$ & $0.62^{+0.08}_{-0.09}$ & $0.59^{+0.10}_{-0.13}$ & $0.73^{+0.05}_{-0.06}$ & $0.44^{+0.12}_{-0.16}$ & $0.85^{+0.06}_{-0.06}$ \\
NFW & HERN & $0.40^{+0.06}_{-0.07}$ & $0.95^{+0.03}_{-0.04}$ & $0.50^{+0.07}_{-0.08}$ & $0.97^{+0.02}_{-0.04}$ & $0.51^{+0.14}_{-0.19}$ & $0.64^{+0.08}_{-0.08}$ & $0.50^{+0.08}_{-0.09}$ & $0.86^{+0.07}_{-0.07}$ & $0.31^{+0.10}_{-0.10}$ & $0.97^{+0.01}_{-0.03}$ \\
HERN & NFW & $0.58^{+0.07}_{-0.08}$ & $0.79^{+0.05}_{-0.05}$ & $0.56^{+0.09}_{-0.11}$ & $0.85^{+0.04}_{-0.05}$ & $0.56^{+0.12}_{-0.17}$ & $0.64^{+0.09}_{-0.10}$ & $0.62^{+0.08}_{-0.10}$ & $0.77^{+0.06}_{-0.06}$ & $0.48^{+0.11}_{-0.13}$ & $0.89^{+0.06}_{-0.07}$ \\
HERN & HERN & $0.48^{+0.05}_{-0.06}$ & $0.96^{+0.02}_{-0.04}$ & $0.57^{+0.06}_{-0.07}$ & $0.97^{+0.01}_{-0.03}$ & $0.54^{+0.13}_{-0.16}$ & $0.66^{+0.10}_{-0.10}$ & $0.55^{+0.07}_{-0.08}$ & $0.90^{+0.06}_{-0.08}$ & $0.42^{+0.08}_{-0.09}$ & $0.98^{+0.01}_{-0.03}$ \\
\midrule
\bottomrule
\end{tabular}
\label{tab.betamajor}
\vspace{-0.1cm}
\tablefoot{We report every combination of mass, surface number density, and velocity anisotropy model, as well as the $1\sigma$ confidence intervals of each value.}
\vspace{-0.4cm}
\end{table*}

\begin{table*}
    \centering
    \caption{Best-fit values of $\beta_0$ and $\beta_\infty$ for the minor galaxy populations.}
    \vspace{-0.2cm}
    \begin{tabular}{cccccccccc}
\toprule
\midrule
\multicolumn{2}{c}{\textbf{gT}} & \multicolumn{2}{c}{Red, high $M_*$} & \multicolumn{2}{c}{Blue, high $M_*$} & \multicolumn{2}{c}{Red, low $M_*$} & \multicolumn{2}{c}{Blue, low $M_*$} \\
\cmidrule(lr){1-2} \cmidrule(lr){3-4} \cmidrule(lr){5-6} \cmidrule(lr){7-8} \cmidrule(lr){9-10} $M(r)$ & $N(R)$ & $\beta_0$ & $\beta_\infty$ & $\beta_0$ & $\beta_\infty$ & $\beta_0$ & $\beta_\infty$ & $\beta_0$ & $\beta_\infty$ \\
\midrule
NFW & NFW & $0.57^{+0.13}_{-0.18}$ & $0.77^{+0.10}_{-0.11}$ & $0.52^{+0.17}_{-0.26}$ & $0.75^{+0.13}_{-0.13}$ & $0.42^{+0.14}_{-0.17}$ & $0.97^{+0.02}_{-0.04}$ & $0.27^{+0.33}_{-0.55}$ & $0.66^{+0.15}_{-0.17}$ \\
NFW & HERN & $0.36^{+0.16}_{-0.23}$ & $0.96^{+0.02}_{-0.07}$ & $0.46^{+0.20}_{-0.27}$ & $0.83^{+0.13}_{-0.16}$ & $0.43^{+0.10}_{-0.13}$ & $0.98^{+0.01}_{-0.02}$ & $0.13^{+0.41}_{-0.59}$ & $0.73^{+0.16}_{-0.18}$ \\
HERN & NFW & $0.43^{+0.19}_{-0.28}$ & $0.86^{+0.09}_{-0.14}$ & $0.46^{+0.20}_{-0.27}$ & $0.85^{+0.11}_{-0.17}$ & $0.50^{+0.11}_{-0.15}$ & $0.97^{+0.01}_{-0.03}$ & $0.01^{+0.49}_{-0.52}$ & $0.84^{+0.13}_{-0.25}$ \\
HERN & HERN & $0.50^{+0.11}_{-0.16}$ & $0.97^{+0.02}_{-0.06}$ & $0.49^{+0.17}_{-0.22}$ & $0.86^{+0.10}_{-0.16}$ & $0.54^{+0.07}_{-0.08}$ & $0.98^{+0.01}_{-0.02}$ & $0.08^{+0.45}_{-0.47}$ & $0.82^{+0.15}_{-0.26}$ \\
\midrule
\midrule
\multicolumn{2}{c}{\textbf{gOM}} & \multicolumn{2}{c}{Red, high $M_*$} & \multicolumn{2}{c}{Blue, high $M_*$} & \multicolumn{2}{c}{Red, low $M_*$} & \multicolumn{2}{c}{Blue, low $M_*$} \\
\cmidrule(lr){1-2} \cmidrule(lr){3-4} \cmidrule(lr){5-6} \cmidrule(lr){7-8} \cmidrule(lr){9-10} $M(r)$ & $N(R)$ & $\beta_0$ & $\beta_\infty$ & $\beta_0$ & $\beta_\infty$ & $\beta_0$ & $\beta_\infty$ & $\beta_0$ & $\beta_\infty$ \\
\midrule
NFW & NFW & $0.60^{+0.11}_{-0.15}$ & $0.73^{+0.08}_{-0.09}$ & $0.59^{+0.12}_{-0.19}$ & $0.69^{+0.09}_{-0.11}$ & $0.46^{+0.13}_{-0.17}$ & $0.92^{+0.04}_{-0.06}$ & $0.36^{+0.26}_{-0.48}$ & $0.59^{+0.09}_{-0.10}$ \\
NFW & HERN & $0.50^{+0.10}_{-0.13}$ & $0.93^{+0.05}_{-0.09}$ & $0.57^{+0.13}_{-0.20}$ & $0.73^{+0.10}_{-0.11}$ & $0.46^{+0.09}_{-0.11}$ & $0.98^{+0.01}_{-0.02}$ & $0.29^{+0.30}_{-0.55}$ & $0.62^{+0.10}_{-0.11}$ \\
HERN & NFW & $0.56^{+0.13}_{-0.21}$ & $0.76^{+0.09}_{-0.13}$ & $0.60^{+0.11}_{-0.18}$ & $0.73^{+0.10}_{-0.12}$ & $0.53^{+0.11}_{-0.13}$ & $0.96^{+0.02}_{-0.04}$ & $0.33^{+0.28}_{-0.61}$ & $0.65^{+0.14}_{-0.14}$ \\
HERN & HERN & $0.57^{+0.09}_{-0.11}$ & $0.96^{+0.03}_{-0.06}$ & $0.59^{+0.12}_{-0.18}$ & $0.76^{+0.12}_{-0.13}$ & $0.56^{+0.07}_{-0.09}$ & $0.98^{+0.01}_{-0.02}$ & $0.30^{+0.28}_{-0.56}$ & $0.66^{+0.14}_{-0.15}$ \\
\midrule
\bottomrule
    \end{tabular}
    \vspace{-0.1cm}
    \tablefoot{We report every combination of mass, surface number density, and velocity anisotropy model, as well as the $1\sigma$ confidence intervals of each value.}
    \label{tab.betaminor}
\vspace{-0.5cm}
\end{table*}

\vspace{-0.3cm}

\section{JE full results and error dissection plots} \label{app.comparedissect}

\begin{figure*}
\centering
\includegraphics[scale=0.31]{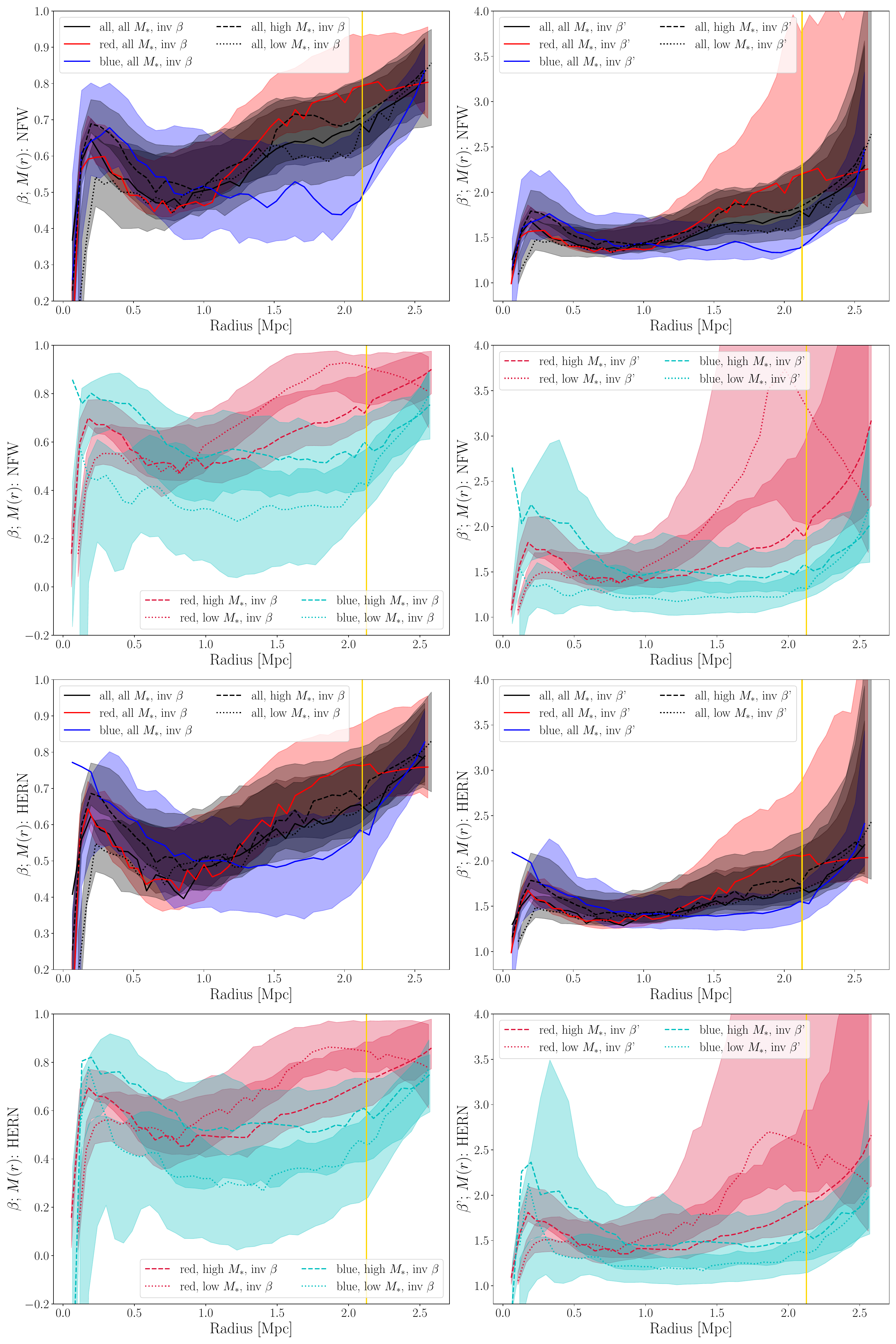}
\caption{Results of the Jeans inversion method for all cluster member populations. In the left column, we show the $\beta(r)$ values, while on the right one we plot those of $\beta'(r)=\sigma_r/\sigma_\vartheta$. In the upper half of this Figure (see label of each vertical axis), results are based on a NFW total mass profile, while in the lower half on Hernquist profile. The shaded regions represent the $1\sigma$ confidence intervals (see Section \ref{subsec.jeanserrs} for further details), while the yellow line represents the $R_\mathrm{200c}$ of the stacked cluster.}
\label{fig.inversiongeneral}
\end{figure*}

\begin{figure*}[h!]
\centering
\includegraphics[scale=0.37]{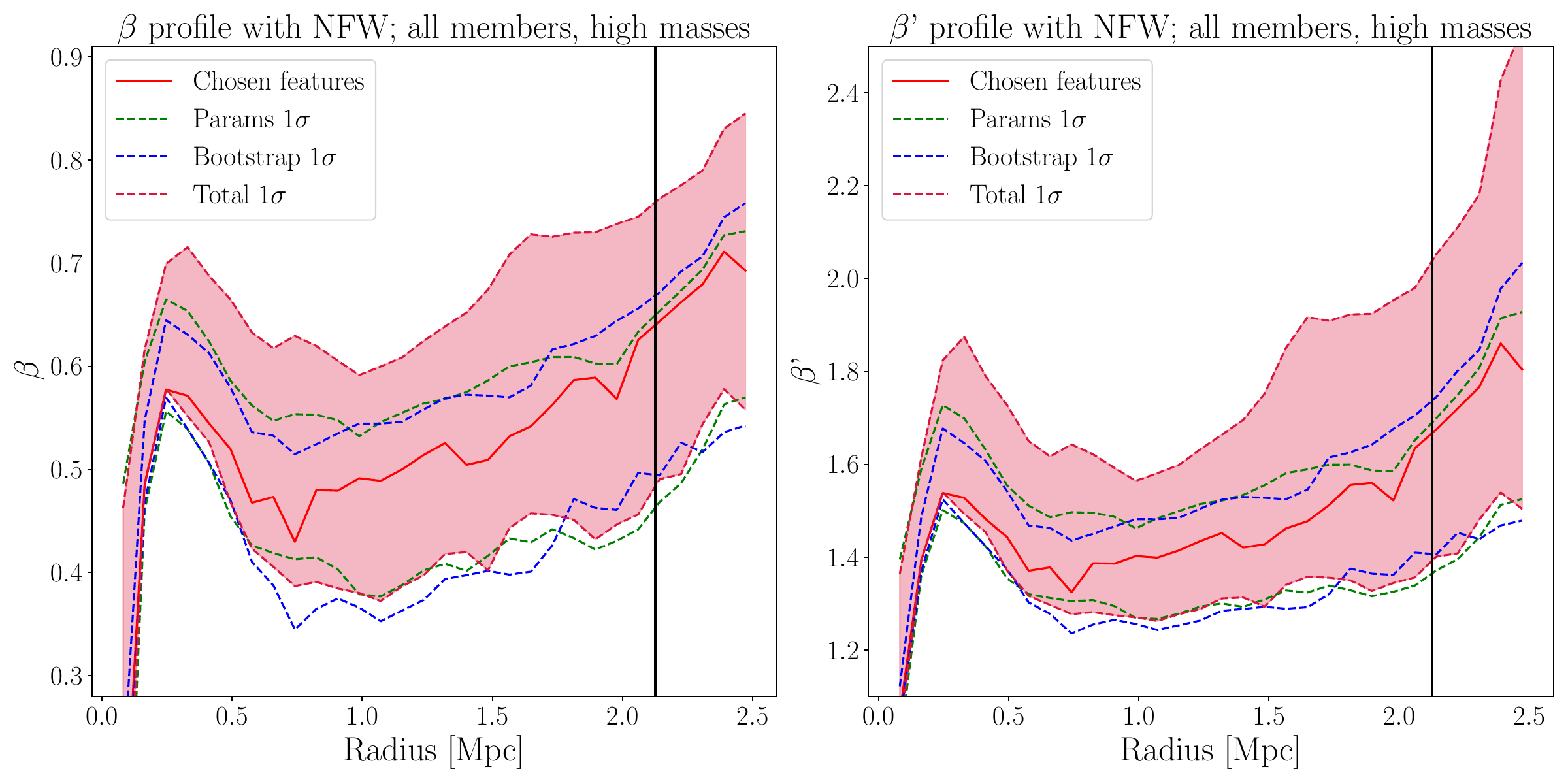}
\vspace{-0.3cm}
\caption{Dissection of the JI confidence intervals for the whole high mass galaxy population (see Section \ref{subsec.jeanserrs}). The green dashed lines represent the systematic uncertainty, while the blue dashed lines represent the uncertainty from bootstrapping both $N(R)$ and $\sigma_\mathrm{los}(R)$. The red dashed lines and the red shaded region mark the overall $1\sigma$ confidence interval. The vertical black line represents the $R_\mathrm{200c}$ of the stacked cluster.}
\label{fig.jeanserr}
\vspace{-0.5cm}
\end{figure*}

In Figure \ref{fig.inversiongeneral} we report the full set of results for the JI analysis.
We analyse the JI total error budget by splitting the systematic and the statistical components.
For the systematics, we compute 250 alternative $\beta(r)$ profiles by varying the values of the set of arbitrarily chosen parameters: we extract from a uniform distribution the values of $\xi$, $f$, $R_\mathrm{inf}$, and the smoothing factors for $N(R)$ and $\sigma_\mathrm{los}(R)$, while we extract the values of the total mass profile scale radius from a Gaussian distribution, centred on the best-fit value of the scale radius and with the error on such scale radius as its standard deviation. For the statistical errors, we generate 250 new $N(R)$ and $\sigma_\mathrm{los}(R)$ profiles through a bootstrap technique, and, consequently, we compute the corresponding $\beta(r)$ profiles. Finally, to evaluate the general (statistical + systematics) confidence intervals, we compute 250 alternative $\beta(r)$ profiles made by picking $N(R)$ and $\sigma_\mathrm{los}(R)$ from the bootstrapped sample and by extracting the arbitrary parameters from their assigned distribution. For all of these sets of $\beta(r)$, we take the interval between the 16th and the 84th percentile of the values of the newly generated $\beta(r)$ at every radius, and we consider it as the $1\sigma$ confidence interval of that kind of error. An example of his operation is shown in Figure \ref{fig.jeanserr}.

\vspace{-0.2cm}

\end{appendix}

\end{document}